\documentclass[
 aps,prb,
 amsmath,amssymb,superscriptaddress,
 reprint,
 a4paper,
] {revtex4-1}

\usepackage{graphicx}
\usepackage{hyperref,amsmath}
\usepackage{braket,amsmath}
\usepackage{dcolumn}
\usepackage{bm}
\usepackage{newfloat}
\usepackage{multibib}
\usepackage{color}

\graphicspath{{./Figs/}}

\begin{document}

\title[]{Silicon quantum processor with robust long-distance qubit couplings}

\author{Guilherme Tosi}
 \affiliation{Centre for Quantum Computation and Communication Technology, School of Electrical Engineering \& Telecommunications, UNSW Sydney, New South Wales 2052, Australia.}
\author{Fahd A. Mohiyaddin}
 \affiliation{Centre for Quantum Computation and Communication Technology, School of Electrical Engineering \& Telecommunications, UNSW Sydney, New South Wales 2052, Australia.}
\author{Vivien Schmitt}
 \affiliation{Centre for Quantum Computation and Communication Technology, School of Electrical Engineering \& Telecommunications, UNSW Sydney, New South Wales 2052, Australia.}
\author{Stefanie Tenberg}
 \affiliation{Centre for Quantum Computation and Communication Technology, School of Electrical Engineering \& Telecommunications, UNSW Sydney, New South Wales 2052, Australia.}
\author{Rajib Rahman}
 \affiliation{Network for Computational Nanotechnology, Purdue University,
 West Lafayette, Indiana 47907, United States}
\author{Gerhard Klimeck}
 \affiliation{Network for Computational Nanotechnology, Purdue University,
 West Lafayette, Indiana 47907, United States}
\author{Andrea Morello}
 \affiliation{Centre for Quantum Computation and Communication Technology, School of Electrical Engineering \& Telecommunications, UNSW Sydney, New South Wales 2052, Australia.}

\date{\today}


\maketitle

\textbf{Practical quantum computers require the construction of a large network of highly coherent qubits, interconnected in a design robust against errors. Donor spins in silicon provide state-of-the-art coherence and quantum gate fidelities, in a physical platform adapted from industrial semiconductor processing. Here we present a scalable design for a silicon quantum processor that does not require precise donor placement and allows hundreds of nanometers inter-qubit distances, therefore facilitating fabrication using current technology. All qubit operations are performed via electrical means on the electron-nuclear spin states of a phosphorus donor. Single-qubit gates use low power electric drive at microwave frequencies, while fast two-qubit gates exploit electric dipole-dipole interactions. Microwave resonators allow for millimeter-distance entanglement and interfacing with photonic links. Sweet spots protect the qubits from charge noise up to second order, implying that all operations can be performed with error rates below quantum error correction thresholds, even without any active noise cancellation technique.}
\vspace{1.5mm}


The successful implementation of quantum algorithms requires incorporation of error correction codes \cite{Terhal2015} that deal with the fragile nature of qubits. The highest tolerances in error rates are found when using nearest-neighbor topological codes \cite{Fowler2012}, long-distance entanglement links \cite{Knill2005} or a combination of both \cite{Nickerson2013}. There exist several physical platforms where state preservation \cite{Maurer2012,Saeedi2013,Muhonen2014}, qubit control \cite{Barends2014,Harty2014,Veldhorst2014,Muhonen2015} and 2-qubit logic gates \cite{Benhelm2008,Barends2014} are achieved with fault-tolerant fidelities. The ultimate goal is to integrate a large number of qubits in expandable arrays to construct a scalable, universal quantum processor.

Donor spin qubits in silicon are an appealing physical platform for that goal, due to their integrability with Metal-Oxide-Semiconductor (MOS) structure and nanometric unit size \cite{Zwanenburg2013}. By using isotopically enriched $^{28}$Si as the substrate material \cite{Itoh2014}, donor spins offer coherence times around a minute \cite{Muhonen2014} or an hour \cite{Saeedi2013}, and control error rates as small as $10^{-4}$ (ref. \onlinecite{Muhonen2015}). However, integrating several of these qubits in a scalable architecture remains a formidable challenge, mainly because of the difficulty in achieving reliable 2-qubit gates.

The seminal Kane proposal \cite{Kane1998} for a nuclear-spin quantum computer in silicon described the use of short-range exchange interactions $J$ between donor-bound electrons, to mediate an effective inter-nuclear coupling of order $\sim 100$~kHz at a $\sim 15$~nm distance. However, the exchange interaction has a exponential and oscillatory spatial behavior that can result in an order of magnitude variation in strength upon displacement by a single lattice site \cite{Koiller2002,Song2016}. Notwithstanding, plenty of progress has been made in the experimental demonstration of the building blocks of a Kane-type processor \cite{Morello2010,Pla2012,Pla2013,Laucht2015}, including the observation of inter-donor exchange \cite{Dehollain2014,Gonzalez2014,Weber2014}. Slightly relaxed requirements on donor placement can be found when using a hyperfine-controlled exchange interaction between electron spin qubits \cite{Kalra2014}, or a slower magnetic dipole-dipole coupling effective at $\sim 30$~nm distances \cite{Hill2015}. Other proposals space donors further apart by introducing some intermediate coupler, $e.g.$ donor chains \cite{Hollenberg2006,Mohiyaddin2016}, charge-coupled devices \cite{Morton2009}, ferromagnets \cite{Trifunovic2013}, probe spins \cite{Ogorman2014} or quantum dots \cite{Pica2015}.

Here we introduce the design of a large-scale, donor-based silicon quantum processor based upon electric dipole interactions. This processor could be fabricated using existing technology, since it does not require precise donor placement. The large inter-qubit spacing, $>150$~nm, leaves sufficient space to intersperse classical control and readout devices, while retaining some of the compactness of atomic-size qubits. New stabilization schemes largely decouple the qubits from electric noise while still keeping them sensitive to electric drive and mutual coupling. Finally, the whole structure retains the standard silicon MOS materials stack, important for ultimate manufacturability.

\ \\
\textbf{Coupling Si:P spin qubits to electric fields}
\vspace{1mm}

\begin{figure}
\centering
\includegraphics[width=\columnwidth]{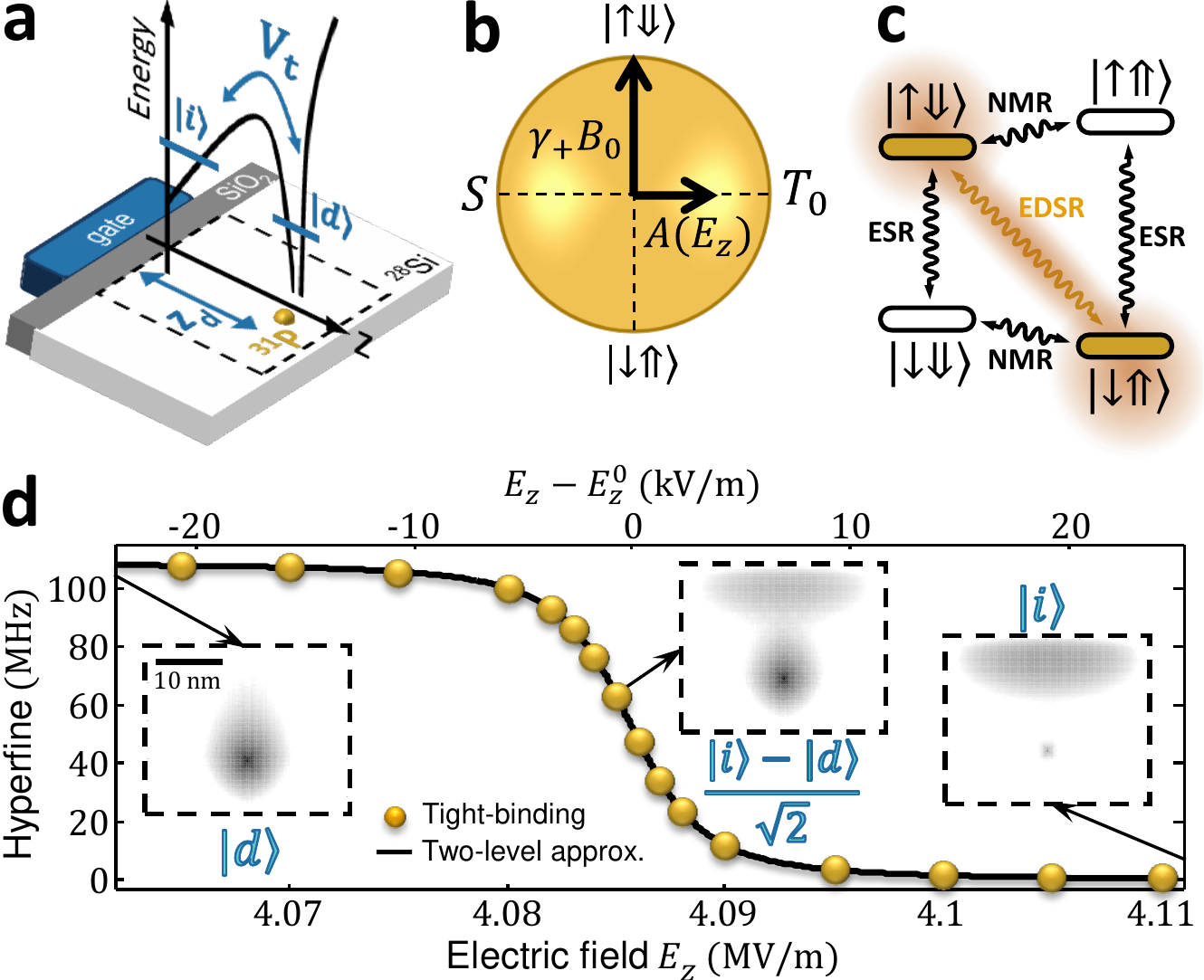}
\caption{\textbf{Coupling Si:P spin qubits to electric fields via hyperfine modulation}. \textbf{a}, Qubit unit cell, in which the electron interface state, $|i\rangle$, is coupled to the donor-bound state, $|d\rangle$, by a tunnel rate $V_t$. Plot shows conduction band profile along $z$. \textbf{b}, Bloch sphere of a flip-flop spin qubit coupled to a vertical electric field $E_z$ via the hyperfine interaction $A$. Singlet and triplet states are denoted by $S=\left(\ket{\downarrow\Uparrow}-\ket{\uparrow\Downarrow}\right)/\sqrt{2}$ and $T_0=\left(\ket{\downarrow\Uparrow}+\ket{\uparrow\Downarrow}\right)/\sqrt{2}$. \textbf{c}, Si:P electron-nuclear spin levels, showing standard electron spin resonance (ESR) and nuclear magnetic resonance (NMR) transitions, together with hyperfine-enabled EDSR. \textbf{d}, Atomistic tight-binding simulations \cite{Klimeck2007} (dots) of the electron-nucleus hyperfine interaction, for a $z_d=15.2$~nm deep donor, as a function of vertical electric field. The solid line is a fit using the simplified two-level Hamiltonian $\mathcal{H}_{\rm orb}+\mathcal{H}_A^{\rm orb}$, which yields $V_t=9.3$~GHz (see Supplementary Information \ref{App:Nemo-orb}). Insets show the electron ground-state, $\ket{g}$, wavefunction, in the region within dashed lines in \textbf{a}, for three different vertical electric fields.}
\label{fig:A(E)}
\end{figure}

\noindent
The phosphorus donor in silicon comprises an electron spin $S = 1/2$ with gyromagnetic ratio $\gamma_e=27.97$~GHz/T and basis states $\ket{\downarrow}, \ket{\uparrow}$, and a nuclear spin $I = 1/2$ with gyromagnetic ratio $\gamma_n=17.23$~MHz/T and basis states $\ket{\Downarrow}, \ket{\Uparrow}$. The electron interacts with the nucleus through the hyperfine coupling $A\approx 117$~MHz. When placed in a large magnetic field $B_0$ ($\gamma_+ B_0 \gg A$, with $\gamma_+=\gamma_e+\gamma_n$), the eigenstates of the system are the separable tensor products of the basis states, i.e. $\ket{\downarrow \Uparrow}, \ket{\downarrow \Downarrow}, \ket{\uparrow \Downarrow}, \ket{\uparrow \Uparrow}$ (Fig. \ref{fig:A(E)}c). The electron and the nucleus can be operated as single qubits by applying oscillating magnetic fields resonant with any of the transitions frequencies between eigenstates that differ by the flipping of one of the spins, e.g. $\ket{\downarrow \Uparrow} \leftrightarrow \ket{\uparrow \Uparrow}$ for the electron qubit, etc (Fig.~\ref{fig:A(E)}c).

We envisage a device where a shallow $^{31}$P donor is embedded in an isotopically pure $^{28}$Si crystal at a depth $z_d$ from the interface with a thin SiO$_2$ layer (Fig. \ref{fig:A(E)}a). The orbital wavefunction $\psi$ of the donor-bound electron can be controlled by a vertical electric field $E_z$ applied by a metal gate on top. It changes from a bulk-like donor state at low electric fields to an interface-like state at high-fields \cite{Calderon2006,Lansbergen2008} (insets in Fig. \ref{fig:A(E)}d). The hyperfine interaction $A(E_z)$, proportional to the square amplitude of the electron wavefunction at the donor site $|\psi(0,0,z_d)|^2$, changes accordingly from the bulk value $A \approx 117$~MHz to $A \approx 0$ when the electron is fully displaced to the interface (Fig.~\ref{fig:A(E)}d). Shifting the electron wavefunction also results in the creation of an electric dipole $\mu_e = ed$, where $e$ is the electron charge and $d$ is the separation between the mean positions of the donor-bound and interface-bound wavefunctions ($d \lesssim z_d$, see Supplementary Information \ref{App:Nemo-orb}). The induced electric dipole $\mu_e$ has been largely overlooked in the past, but plays a crucial role in this proposal.

The key idea is to define a new qubit, called henceforth the \emph{flip-flop qubit}, described in the subspace spanned by the states $\ket{\downarrow \Uparrow}, \ket{\uparrow \Downarrow}$. Transitions between these basis states cannot be induced by magnetic resonance, because there is no change in the $z$-component of the total angular momentum. However, the hyperfine interaction, $A\mathbf{S\cdot I}$,  is a transverse term in the flip-flop basis, since its eigenstates are $S=(\ket{\downarrow \Uparrow} - \ket{\uparrow \Downarrow})/\sqrt{2}$, $T_0=(\ket{\downarrow \Uparrow} + \ket{\uparrow \Downarrow})/\sqrt{2}$ (Fig.~\ref{fig:A(E)}b). Therefore, electrically modulating $A(E_z)$ at the frequency

\begin{equation} \label{eq:e_ff}
\epsilon_{\rm ff}(A)=\sqrt{\left(\gamma_+B_0\right)^2+\left[A\left(E_z\right)\right]^2},
\end{equation}

corresponding to the flip-flop qubit energy splitting, causes an electric dipole spin resonance (EDSR) transition between the $\ket{\downarrow \Uparrow}, \ket{\uparrow \Downarrow}$ basis states \cite{Laird2007,Luo2012} (Fig.~\ref{fig:A(E)}c). This transition is faster at the ``ionization point'', where the electron is shared halfway between donor and interface, since $A(E_z)$ can vary strongly upon the application of a small voltage on the top gate.

\ \\
\textbf{Electrical noise and relaxation}
\vspace{1mm}

\noindent
Since the qubit operation is based upon the use of electric fields, a natural concern is the fragility of the qubit states in the presence of electric noise. Below we show that there are special bias points that render the flip-flop qubit operation highly robust against noise.

A quantum-mechanical description of the system is obtained by treating also the electron position as a two-level system (effectively a charge qubit; see Supplementary Information \ref{App:Nemo-orb} for a justification of this two-level approximation), where the vertical position of the electron is represented by a Pauli $\sigma_z$ operator, with eigenvectors $|d\rangle$, for the electron at the donor, and $|i\rangle$ at the interface (Fig. \ref{fig:A(E)}a,d). The simplified orbital Hamiltonian reads (in units of Hz):

\begin{equation} \label{eq:H_orb}
\mathcal{H}_{\rm orb}=\frac{V_t \sigma_x-\left[e(E_z - E_z^0) d/h\right]\sigma_z}{2},
\end{equation}

where $V_t$ is the tunnel coupling between the donor and the interface potential wells, $E_z^0$ is the vertical electric field at the ionization point and $h$ is the Planck constant. The electron ground $|g\rangle$ and excited $|e\rangle$ orbital eigenstates depend on $E_z$ (Fig.~\ref{fig:A(E)}d) and have an energy difference given by:

\begin{equation} \label{eq:e_o}
\epsilon_{\rm o}=\sqrt{\left(V_t\right)^2+\left[e(E_z - E_z^0) d/h\right]^2}
\end{equation}

\begin{figure*}
\centering
\includegraphics[width=\textwidth]{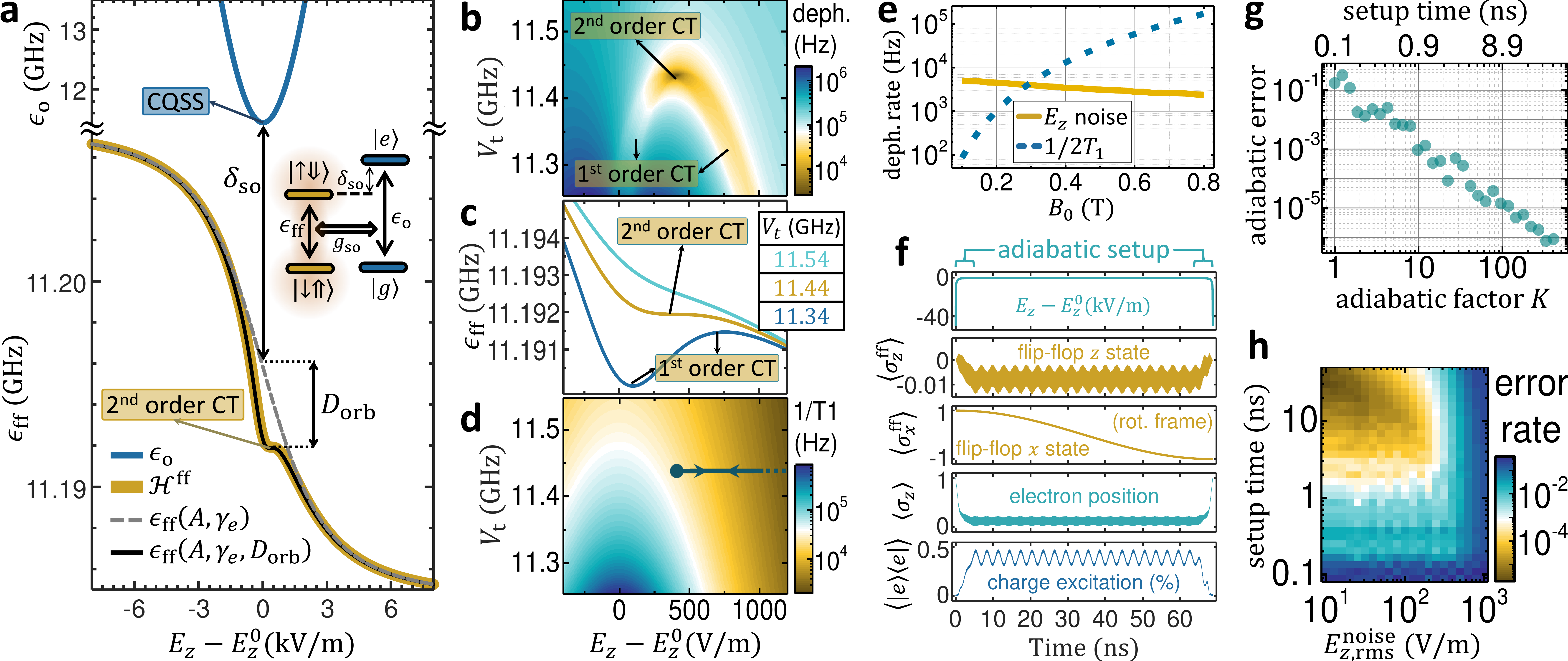}
\caption{\textbf{Robustness to electric noise and high-fidelity adiabatic $z$-gates}. \textbf{a}, Charge, $\epsilon_{\rm o}$, and flip-flop, $\epsilon_{\rm ff}$, qubits transition frequencies as a function of vertical electric field $E_z$, for $B_0=0.4$~T, $A=117$~MHz, $d=15$~nm, $\Delta_\gamma=-0.2\%$ and $V_t=11.44$~GHz. Inset shows the level diagram of flip-flop states coupled to charge states. CT stands for ``clock transition" and CQSS for ``charge qubit sweet spot". \textbf{b}, Estimated flip-flop qubit dephasing rate, assuming electric field noise $E_{z, \rm rms}^{\rm noise}=100$~V/m. \textbf{c}, $E_z$-dependence of flip-flop precession frequency for the three indicated tunnel coupling values. \textbf{d}, Flip-flop qubit relaxation rate, with arrows indicating adiabatic path used for $z$-gates. \textbf{e}, Flip-flop qubit dephasing rate due to $E_z$ noise and relaxation, at $2^{\rm nd}$-order CTs for each $B_0$. \textbf{f}, Time-evolution of an adiabatic  ($K=50$) $\pi$ $z$-gate on state $\ket{g}\otimes(\ket{\downarrow\Uparrow}+\ket{\uparrow\Downarrow})/\sqrt{2}$, showing applied electric field and flip-flop/charge states. Outer brackets denote the expected value of an operator. $\sigma_z^{\rm ff}=\ket{\uparrow\Downarrow}\bra{\uparrow\Downarrow}-\ket{\downarrow\Uparrow}\bra{\downarrow\Uparrow}$ and $\sigma_x^{\rm ff}=\ket{+_x^{\rm ff}}\bra{+_x^{\rm ff}}-\ket{-_x^{\rm ff}}\bra{-_x^{\rm ff}}$, where $\ket{+_x^{\rm ff}}=\left(\ket{\uparrow\Downarrow}+\exp{\left(-i2\pi\epsilon_{\rm ff}^{t=0}\right)}\ket{\downarrow\Uparrow}\right)/\sqrt{2}$ and $\ket{-_x^{\rm ff}}=\left(\ket{\uparrow\Downarrow}+\exp{\left(-i2\pi\epsilon_{\rm ff}^{t=0}-i\pi\right)}\ket{\downarrow\Uparrow}\right)/\sqrt{2}$. Fast oscillations between the charge and flip-flop states are due to small deviations from perfect adiabaticity. \textbf{g}, $\pi$ $z$-gate leakage error for different adiabatic setup times, which are set by the factor $K$. \textbf{h}, $\pi$ $z$-gate error due to quasi-static $E_z$ noise, at the $2^{\rm nd}$-order CT at $B_0=0.4$~T, for different noise amplitudes and adiabatic setup times.}
\label{fig:clock}
\end{figure*}

At the ionization point, the energy difference between eigenstates $|e\rangle=(|d\rangle+|i\rangle)/\sqrt{2}$ and $|g\rangle=(|d\rangle-|i\rangle)/\sqrt{2}$ is minimum and equal to $V_t$ (Fig. \ref{fig:clock}a), and therefore first-order insensitive to electric noise, $\partial\epsilon_{\rm o}/\partial E_z=0$. This bias point is referred to as the ``charge qubit sweet spot''\cite{Kim2015a} (CQSS -- Fig. \ref{fig:clock}a).

Conversely, the bare flip-flop qubit energy is expected to depend strongly on $E_z$, through the combined effect of the hyperfine interaction $A$ (Eq. \ref{eq:e_ff}) and the orbital dependence of the electron gyromagnetic ratio, $\gamma_e$. Indeed, the gyromagnetic ratio of an electron confined at a Si/SiO$_2$ interface can differ from that of a donor-bound electron by a relative amount $\Delta_{\gamma}$ up to 0.7\% \cite{Rahman2009a}. Therefore, the Zeeman terms in the Hamiltonian must include a dependence of the electron Zeeman splitting on its orbital position, i.e. the charge qubit $\sigma_z$ operator:

\begin{equation} \label{eq:H_Zeeman_orb}
\mathcal{H}_{B_0}^{\rm orb}=\gamma_e B_0\left[1+\left(\frac{1+\sigma_z}{2}\right)\Delta_\gamma\right]S_z - \gamma_n B_0 I_z.
\end{equation}

We can also write the hyperfine coupling as an operator that depends on the charge qubit state:

\begin{equation} \label{eq:H_A}
\mathcal{H}_{A}^{\rm orb}=A\left(\frac{1 - \sigma_z}{2}\right){\bf S\cdot I}
\end{equation}

Indeed, this simple two-level approximation, shown as a black line in Fig~\ref{fig:A(E)}d, reproduces the full tight-biding simulations (yellow dots).

The overall flip-flop qubit transition frequency as a function of $E_z$ becomes:

\begin{equation} \label{eq:e_ff_ge}
\epsilon_{\rm ff}(A,\gamma_e)=\sqrt{\left[\gamma_e(E_z)+\gamma_n\right]^2{B_0}^2+\left[A\left(E_z\right)\right]^2},
\end{equation}

shown in Fig.~\ref{fig:clock}a (dashed line), where we assumed $\Delta_\gamma=-0.2\%$. \cite{Rahman2009a}. $\epsilon_{\rm ff}(A,\gamma_e)$ shows a steep slope around the ionization point, mostly caused by the $E_z$-dependence of $\gamma_e$ (the dependence on $A$ is less significant because $\gamma_+B_0\gg A$). Therefore, while $E_z\approx E_z^0$ is the fastest operation point for the flip-flop qubit driven by a resonant modulation of $A$, it can also be the most prone to qubit dephasing from charge and gate noise, through the influence of $E_z$ on $\gamma_e$.

However, computing instead the \emph{full} flip-flop qubit Hamiltonian,

\begin{equation} \label{eq:H_ff_ge}
\mathcal{H}_{\rm ff} = \mathcal{H}_{B_0}^{\rm orb} +\mathcal{H}_{A}^{\rm orb} + \mathcal{H}_{\rm orb},
\end{equation}

reveals that the qubit transition frequency has an extra bend around the ionization point (Fig.~\ref{fig:clock}a -- thick yellow line). This comes from Eq. \ref{eq:H_A}, which provides a transverse coupling $g_{\rm so}$ between the flip-flop and charge qubits (inset in Fig. \ref{fig:clock}a):

\begin{equation} \label{eq:g_so}
g_{\rm so}=\frac{A}{4}\frac{V_t}{\epsilon_{\rm o}}
\end{equation}

As a result, the electron orbit dispersively shifts the flip-flop qubit by, to second order:

\begin{equation} \label{eq:Dorb}
D_{\rm orb}(E_z)=\frac{[g_{\rm so}(E_z)]^2}{\delta_{\rm so}(E_z)},
\end{equation}

where $\delta_{\rm so}=\epsilon_{\rm o} - \epsilon_{\rm ff}$, reducing the flip-flop qubit frequency to:

\begin{equation} \label{eq:e_ff_ge_Dorb}
\epsilon_{\rm ff}(A,\gamma_e,D_{\rm orb})=\epsilon_{\rm ff}(A,\gamma_e)-D_{\rm orb}(E_z),
\end{equation}

$D_{\rm orb}(E_z)$ is largest around $E_z\approx E_z^0$, since $\delta_{\rm so}$ is lowest (i.e. the charge qubit frequency comes closest to the flip-flop qubit, Fig. \ref{fig:clock}a) and $g_{\rm so}$ is highest. Eq.~\ref{eq:e_ff_ge_Dorb} (thin black line in Fig. \ref{fig:clock}a) agrees with full numerical simulations of the Hamiltonian in Eq. \ref{eq:H_ff_ge}. 

Such a dispersive shift stabilizes the flip-flop precession frequency against noise. To quantify that, we assume a quasi-static electric field noise with 100 V/m r.m.s. amplitude along the donor-dot direction ($z$-axis in Fig.~\ref{fig:A(E)}a). This noise is equivalent to a $1.5~\mu$eV charge detuning noise for $d=15$~nm, consistent with measured values \cite{Freeman2016,Thorgrimsson2016,Harvey-Collard2015} -- see Supplementary Information \ref{App:Elec_noise}. The estimated -- see Methods section -- dephasing rates can be as low as $1/T_2^{\ast} \approx 3$~kHz (Fig. \ref{fig:clock}b), comparable to the ones due to magnetic noise ($1/T_2^{\ast} \approx 1$~kHz in $^{28}$Si nanostructures \cite{Muhonen2014}). This can be understood from Fig.~\ref{fig:clock}c, which shows the qubit precession frequency dependence on $E_z$, for three different values of $V_t$. For small detunings $\delta_{\rm so}$, $i.e.$ $V_t$ close to $\epsilon_{\rm ff}$, the dispersive shift around the ionization point is strong, yielding two first-order ``clock transitions'' (CT), where $\partial\epsilon_{\rm ff}/\partial E_z=0$ where the dephasing rate is reduced. By increasing $V_t$, the two first-order points merge into a single one in which both the first and second derivatives vanish, yielding the slowest qubit dephasing.

Another source of errors could come from relaxation via coupling to phonons. This is not an issue for bulk donors, where electron spin relaxation time is $T_{1,\rm s}\gg 1$~s \cite{Morello2010}. However, due to the particular valley composition of the flip-flop qubit near the ionization point, its relaxation rate ${1}/{T_{1,\rm ff}}$ due to charge-phonon coupling is enhanced \cite{Boross2016}. We estimate it by noting that, if $\delta_{\rm so}\gg g_{\rm so}$, ${1}/{T_{1,\rm ff}}$ is equal to the amount of charge excited state in the flip-flop eigenstates \cite{Blais2004} times the charge relaxation rate \cite{Boross2016}:

\begin{subequations}
\begin{equation}\label{eq:T1ff}
{1}/{T_{1,\rm ff}}=\left({g_{\rm so}}/{\delta_{\rm so}}\right)^2/{T_{1,\rm o}},
\end{equation}
\begin{equation}\label{eq:T1o}
{1}/{T_{1,\rm o}}=\Theta\epsilon_{\rm o}{V_t}^2,
\end{equation}
\end{subequations}

where $T_{1,\rm o}$ is the charge qubit lifetime and $\Theta\approx2.37\times10^{-24}~{\rm s}^2$ is determined by the silicon crystal properties \cite{Boross2016}. Therefore, as can be seen from Fig. \ref{fig:clock}d, the higher the detuning $\delta_{\rm so}$, the slower the relaxation. In particular, at the $2^{\rm nd}$-order CT, the qubit dephasing can be limited by relaxation, $1/T_2^*=1/2T_1\approx10^4$~Hz. This limitation can be overcome by reducing $B_0$ (Fig. \ref{fig:clock}e).

Tuning a flip-flop qubit at a clock transition requires the ability to tune the tunnel coupling $V_t$. The latter is difficult to control at the fabrication stage, given its exponential dependence on donor depth, together with oscillations at the atomic scale \cite{Calderon2008} arising from a similar valley interference effect as the one afflicting the exchange interaction \cite{Koiller2002}. To overcome that, $V_t$ can be electrostatically tuned, by at least 2 orders of magnitude, by using a gate stack identical to the well-established scheme for the confinement of single electrons in Si quantum dots \cite{Veldhorst2014} -- see Supplementary Information \ref{App:Vt-tunability}.

The presence of slow dephasing regions is important to control the qubit phase with high fidelity. In our quantum processor, idle qubits are decoupled from electric fields by fully displacing the electron either to the interface or to the donor. Performing quantum operations on the qubit requires displacing the electrons close to the ionization point, which in turn changes its precession frequency (Fig.~\ref{fig:clock}a). As a result, the accumulated phase must be corrected after quantum operations. This is optimally done by moving the electron to the $2^{\rm nd}$-order clock transition, therefore minimizing dephasing errors. At this point, the flip-flop qubit phase precesses $\sim\Delta_\gamma\gamma_e B_0/2-D_{\rm orb}$ faster than its idle point, and therefore any phase correction in a $2\pi$ period can be applied within tens of ns. The dephasing rate at the CT, on the order of a few kHz, would cause very small errors ($<10^{-4}$). However, while moving the electron from the interface towards the donor, the flip-flop qubit goes through regions of fast dephasing (Fig.~\ref{fig:clock}b), and therefore this operation has to be performed as quickly as possible. It also has to be slow enough as to avoid erros due to non-adiabaticity, which include \textit{e.g.} leakage to unwanted high-energy states. These errors depend on the adiabatic factor $K$, which quantifies the fractional rate of change of the system's eigenstates (the higher the value of $K$, the more adiabatic and slower is the process -- see Methods).

In Fig.~\ref{fig:clock}f we plot the time dynamics of an initial state $\ket{g}\otimes(\ket{\downarrow\Uparrow}+\ket{\uparrow\Downarrow})/\sqrt{2}$ while sweeping $E_z$ adiabatically ($K=50$) to move the electron from the interface to the $2^{\rm nd}$-order CT and back, in order to realize a $\pi$ $z$-gate. The initial adiabatic setup part consists of a fast sweep (0.8~ns), allowed by the large charge qubit splitting when $E_z \gg E_z^0$, followed by a slower sweep (3.5~ns), limited by the proximity of excited charge states to the flip-flop qubit when $E_z \approx E_z^0$. The electron then remains at the CT for 60~ns, before adiabatically moving back to the interface. During the total 69~ns, the flip-flop qubit phase is shifted by $\pi$, with adiabatic errors, averaged over a set of initial flip-flop states -- see Methods -- around $10^{-4}$. These errors can be controlled with the factor $K$, which sets the setup time (see Fig.~\ref{fig:clock}g).

Quasi-static $E_z$ noise increases errors, due to dephasing (Fig. \ref{fig:clock}h). At realistic noise levels (100~V/m), the gate error rate is found to be $<10^{-4}$. Similar error levels arise due to relaxation, which remains below $3\times10^4$~Hz (Fig.~\ref{fig:clock}d).

Note that the presence of clock transitions does not affect the ability to use $E_{\rm ac}$ to resonantly drive the qubit, since the transverse term $A(E_z)$ still responds fully to the electric field (this is similar to the case of magnetic clock transitions, e.g. in Si:Bi \cite{Wolfowicz2013}).

\ \\
\textbf{Electric drive of flip-flop qubit}
\vspace{1mm}

\noindent

\begin{figure}
\centering
\includegraphics[width=\columnwidth]{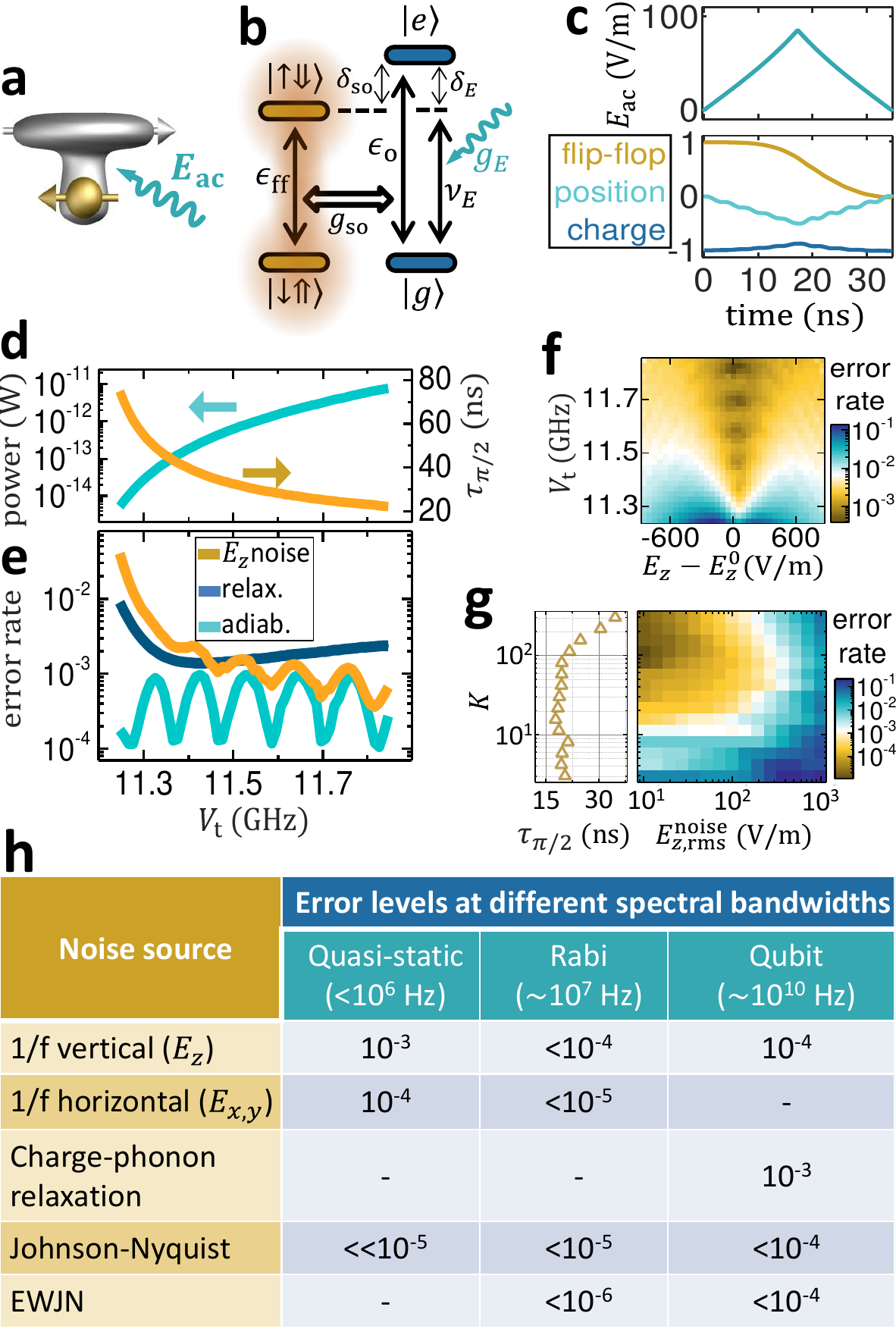}
\caption{\textbf{High-fidelity electrically-driven adiabatic 1-qubit x(y)-gates}. \textbf{a}, Spatial representation and \textbf{b}, level diagram, for electrical drive of a flip-flop qubit, showing partially ionized electron wavefunction and spin arrows. \textbf{c}, Time-dependent adiabatic drive amplitude and qubit dynamics of a $\pi/2$ $x$-gate, for $K=30$, $B_0=0.4$~T, $E_z=E_z^0$ and $V_t=11.5$~GHz. Bottom plot shows flip-flop z state, $\langle\sigma_z^{\rm ff}\rangle$, electron position, $\langle\sigma_z\rangle$, and charge qubit state, $\langle\ket{e}\bra{e}-\ket{g}\bra{g}\rangle$. For the same parameters, \textbf{d} shows the averaged drive power and gate time, and \textbf{e} the error rates for different $V_t$. To estimate the drive power, we assumed a 50~$\Omega$ line in which a $1~{\rm \mu V}$ AC voltage produces a 10~V/m AC vertical electric field. \textbf{f}, Estimated flip-flop qubit $\pi/2$ $x$-gate error due to quasi-static noise with amplitude $E_{z, \rm rms}^{\rm noise}=100$~V/m. \textbf{g}, Dependence of gate error rate on the electric noise r.m.s. amplitude and adiabatic factor K (which sets the gate time). \textbf{h}, Estimated gate error rates from different noise sources, according to Supplementary Information \ref{App:Elec_noise}. Hyphens indicate inexistent/negligible errors.}
\label{fig:1-qubit}
\end{figure}

We now explain how high-fidelity 1-qubit $x(y)$-gates can be achieved via electric drive of the flip-flop qubit. The fastest 1-qubit gates are obtained when the electron is around the ionization point, where $\partial A/\partial E_z$ is maximum (Fig. \ref{fig:A(E)}d). A vertical oscillating electric field of amplitude $E_{\rm ac}$ is applied (Fig. \ref{fig:1-qubit}a) in resonance with the flip-flop qubit, \textit{i.e}, $\nu_E=\epsilon_{\rm ff}$. A large detuning $\delta_{\rm so}\gg g_{\rm so}$ ensures the least amount of the charge excited state $\ket{e}$ in the qubit eigenstates, minimizing qubit relaxation via charge-phonon coupling. The flip-flop qubit is still driven, via a second-order process, at a rate (half-Rabi frequency):

\begin{equation} \label{eq:g_E_ff}
g^{\rm ff}_{E}=\frac{g_{\rm so}g_{E}}{2}\left(\frac{1}{\delta_{\rm so}}+\frac{1}{\delta_E}\right),
\end{equation}

where $\delta_E=\nu_E-\epsilon_{\rm o}$ and $g_{E}$ is the driven electric coupling rate between the two charge eigenstates:

\begin{equation} \label{eq:g_E}
g_{E}=\frac{e E_{\rm ac} d}{4h}\frac{V_t}{\epsilon_{\rm o}},
\end{equation}

where $E_{\rm ac}$ is the amplitude of a sinusoidal drive. Equation \ref{eq:g_E_ff} provides another explanation why the fastest 1-qubit gates are obtained when the electron is at the ionization point: $\delta_{\rm so}$ and $\delta_E$ are minimum ($\epsilon_{\rm o}$ is minimum), and $g_{\rm so}$ and $g_E$ are maximum (Eqs. \ref{eq:g_so} and \ref{eq:g_E}).

The electrical drive can cause some excitation of the charge qubit. It is therefore convenient to turn $E_{\rm ac}$ on/off adiabatically to make sure the charge is de-excited at the end of the gate. Figure \ref{fig:1-qubit}c shows the $E_{\rm ac}$ time evolution needed for a $\pi/2$ $x$-gate, where we have assumed an adiabatic factor $K=30$, sufficient for leakage errors $<10^{-3}$. $E_{\rm ac}$ increases steadily until a $\pi/4$ rotation is completed, after which $E_{\rm ac}$ is gradually switched off to achieve an adiabatic $\pi/2$ $x$-gate. An average 4\% excitation of the charge qubit causes a $\sim4\times10^4$~Hz relaxation rate of the encoded quantum state (Eq. \ref{eq:T1o}), or error levels close to $10^{-3}$. 

We then investigate how the total $\pi/2$ $x$-gate errors depend on the biasing of the electron wavefunction. At the ionization point, $E_z=E_z^0$, error levels close to $10^{-3}$ are found over a wide range of $V_t$ (Fig. \ref{fig:1-qubit}e). The $K=30$ choice ensures adiabatic errors $<10^{-3}$ with an oscillatory character typical of adiabatic processes \cite{Oh2013}. At small $V_t$ (and therefore small detuning $\delta_{\rm so}$), the qubit eigenstates contain a substantial amount of charge, causing more errors due to charge-phonon relaxation. Increasing the detuning $\delta_E$ with larger $V_t$ allows for a faster adiabatic sweep and higher powers (Fig.~\ref{fig:1-qubit}d), yielding shorter gate times and therefore less errors due to quasi-static noise. Still, the incident power is at least three orders of magnitude lower than the one needed to drive donor electron spin qubits, at the same Rabi frequency, with oscillating magnetic fields \cite{Pla2012,Muhonen2014}.

As Fig. \ref{fig:1-qubit}f shows, low error rates are still available away from the ionization point, even though best values are found at $E_z=E_z^0$. This is because our gate times are so fast that dephasing, and therefore CT's, do not play a crucial role. Instead, quasi-static $E_z$ noise cause errors mainly by modulating the driving strength $g_E^{\rm ff}$, causing ``gate time jitter''. Indeed, the gate time is sensitive to the orbital transition frequency $\epsilon_{\rm o}$ (Eq. \ref{eq:g_E_ff}), and therefore gate errors are minimized close to the charge qubit sweet spot (CQSS), where $\partial\epsilon_{\rm o}/\partial E_z=0$ (Fig. \ref{fig:clock}a).

Finally, as Fig. \ref{fig:1-qubit}g shows, lower quasi-static $E_z$ noise can cause less errors, provided that the adiabatic factor $K$ is increased, to reduce leakage errors, up to an optimum value where gate times are still fast as to keep noise errors low. Relaxation errors could also be reduced by reducing $B_0$ (recall Fig.~\ref{fig:clock}e).

A number of other noise sources, including high frequency charge noise, Johnson-Nyquist and evanescent-wave Johnson noise \cite{Henkel1999} (EWJN) also affect qubits that are sensitive to electric fields. However, as we discuss in Supplementary Information \ref{App:Elec_noise}, the corresponding error rates are much lower than the ones already previously mentioned -- see all estimated error levels in Fig.~\ref{fig:1-qubit}h.

\ \\
\textbf{Two-qubit coupling via electric dipole interaction}
\vspace{1mm}

\begin{figure*}
\centering
\includegraphics[width=\textwidth]{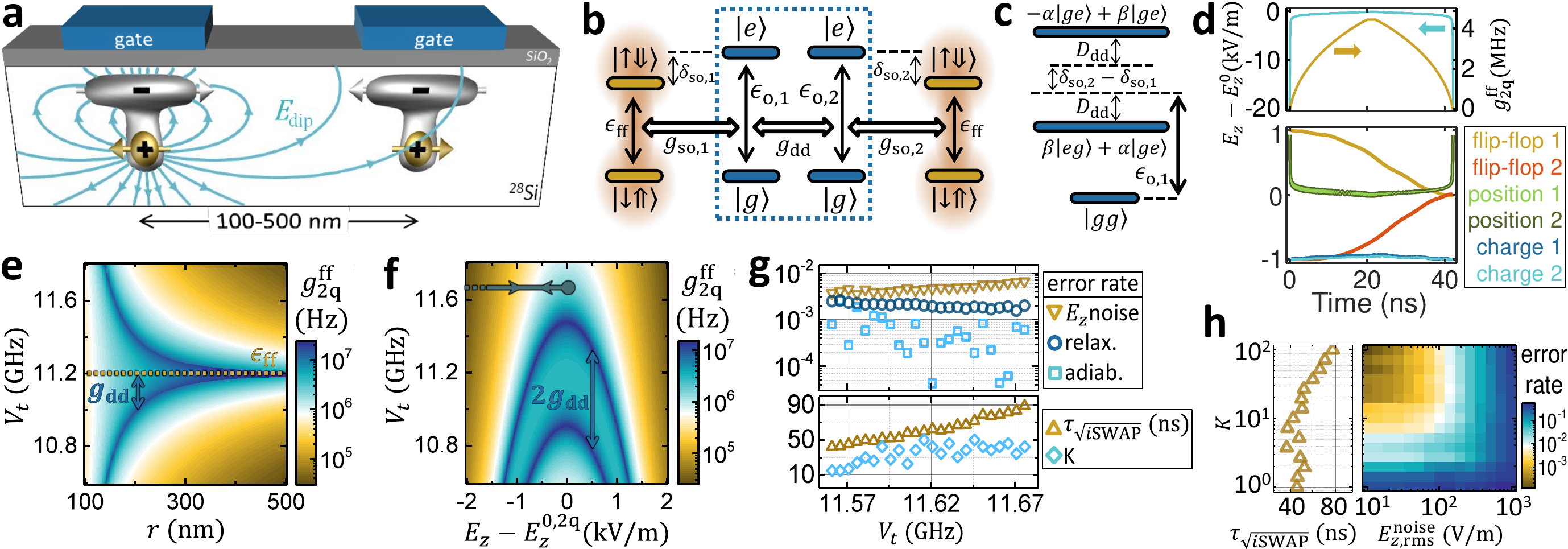}
\caption{\textbf{High-fidelity adiabatic $\sqrt{i{\rm SWAP}}$ gates between two distant flip-flop qubits via electric dipole-dipole interactions}.
\textbf{a}, Device scheme for coupling qubits, showing dipole field lines, $\textit{\textbf{E}}_{\rm dip}$, produced by the dipole on the left.
\textbf{b}, Level diagram for two-qubit coupling via direct dipole-dipole interaction.
\textbf{c}, Lowest molecular eigenstates for the two charge qubits inside dashed rectangle in \textbf{b}. Eigenenergy shift equals $D_{\rm dd}=(\delta_{\rm so,2}-\delta_{\rm so,1})\left(1+[2g_{\rm dd}/(\delta_{\rm so,2}-\delta_{\rm so,1})]^2\right)/2$. Eigenstate coefficients are $\beta=\theta/\sqrt{\theta^2+1}$ and $\alpha=\phi/\sqrt{\phi^2+1}$, with $\theta,\phi=[(\delta_{\rm so,2}-\delta_{\rm so,1})\pm\sqrt{(\delta_{\rm so,2}-\delta_{\rm so,1})^2+(2g_{\rm dd})^2}]/(2g_{\rm dd})$.
\textbf{d}, Time evolution of an adiabatic $\sqrt{i \rm SWAP}$ gate, for $K=30$, $r=180$~nm, $B_0=0.4$~T and $V_t=11.58$~GHz.
Effective coupling between 2 flip-flop qubits as a function of $V_{t,1}=V_{t,2}=V_t$, interdistance $r$ (\textbf{e}) and electric field $E_{z,1}=E_{z,2}=E_z$ (\textbf{f}). Upper arrows in \textbf{f} represent adiabatic path followed for 2-qubit gates. $E_z^{\rm 0,2q}$ is the ionization point in the presence of a second qubit, $E_z^{\rm 0,2q}=E_z^0-2g_{\rm dd}h/(2eL_i)$.
\textbf{g}, Optimized $\sqrt{i{\rm SWAP}}$ gate errors, time and adiabatic factor $K$.
\textbf{h}, Optimized error rate due to quasi-static $E_z$-noise for different noise amplitudes and adiabatic factor $K$ (which sets the gate time).}
\label{fig:2-qubit}
\end{figure*}

\noindent
We now present the new method to couple donor spins that lies at the heart of our scalable quantum processor. It exploits the electric dipole that naturally arises when a donor-electron wavefunction is biased to the ionization point (Fig.~\ref{fig:2-qubit}a), due to the fact that a negative charge has been partly displaced away from the positive $^{31}$P nucleus. The electric field produced by this induced dipole can, in turn, introduce a coupling term in a nearby donor which is also biased at the ionization point.

The interaction energy between two distant dipoles, $\mu_1$ and $\mu_2$, oriented perpendicularly to their separation, $r$, is \cite{Ravets2014} $V_{\rm dip}=\mu_1\mu_2/(4\pi\varepsilon_r\varepsilon_0r^3)$, where $\varepsilon_0$ is the vacuum permittivity and $\varepsilon_r$ the material's dielectric constant ($\varepsilon_r=11.7$ in silicon). The electric dipole of each donor-interface state is $\mu_i=ed_i(1+\sigma_{z,i})/2$, implying that the dipole-dipole interaction Hamiltonian is:

\begin{subequations}
\begin{equation} \label{eq:H_dipdip}
\mathcal{H}_{\rm dip}=V_{\rm dd}\left(\sigma_{z,1}\sigma_{z,2}+\sigma_{z,1}+\sigma_{z,2}\right)
\end{equation}
\begin{equation} \label{eq:V_dd}
V_{\rm dd}=\frac{1}{16\pi\varepsilon_0\varepsilon_r h}\frac{ed_1~ed_2}{r^3}
\end{equation}
\end{subequations}

This electric dipole-dipole interaction is therefore equivalent to a small shift in the equilibrium orbital position of both electrons plus a coupling term between the charge qubits (blue dashed rectangle in Fig. \ref{fig:2-qubit}b) equal to:

\begin{equation} \label{eq:g_dd}
g_{\rm dd}=V_{\rm dd}\frac{V_{t,1}V_{t,2}}{\epsilon_{\rm o,1}\epsilon_{\rm o,2}}
\end{equation}

Most importantly, since each flip-flop qubit is coupled to their electron position (Eq.~\ref{eq:H_A}), the electric dipole-dipole interaction provides a natural way to couple two distant flip-flop qubits.

Indeed, the effective coupling rate between two flip-flop qubits at the ionization point, Fig.~\ref{fig:2-qubit}e, exceeds $10^7$~Hz around two narrow regions. These bands can be understood from the energy-level diagram shown in Fig.~\ref{fig:2-qubit}c. The two charge qubits in Fig.~\ref{fig:2-qubit}b form hybridized molecular states, which are coupled to each flip-flop qubit. The 2-qubit coupling rate is maximum when in resonance with a molecular state. However, this regime induces too many relaxation errors due to resonant charge excitation. Therefore it is best to detune the flip-flop qubits from the molecular states, while still keeping a substantial inter-qubit coupling rate, via a second-order process, equal to:

\begin{equation}\label{eq:flipdipSWAP_Delta}
g_{\rm 2q}^{\rm ff}=g_{\rm so,1}g_{\rm so,2}\alpha\beta\left(\frac{1}{D_{\rm dd}-\delta_{\rm so,1}}+\frac{1}{D_{\rm dd}+\delta_{\rm so,2}}\right),
\end{equation}

where $D_{\rm dd}$ is the charge eigenenergies shift and $\alpha$, $\beta$ the eigenstates coefficients -- see Fig. \ref{fig:2-qubit}c caption.

2-qubit gates start with both electrons at the interface, where qubits are decoupled since the electric dipoles and the hyperfine interactions are first-order insensitive to vertical electric fields. Indeed, from Eq.~\ref{eq:flipdipSWAP_Delta}, $g_{\rm 2q}^{\rm ff}$ is negligible since $g_{\rm so}$ vanishes and $\delta_{\rm so}$ diverges. The electrons are then simultaneously and adiabatically displaced to the ionization point for a time necessary for an $\sqrt{i\mathrm{SWAP}}$ gate, before returning to the interface. In Fig.~\ref{fig:2-qubit}d we show the dynamics of a 2-qubit gate performed with an adiabatic factor $K=30$, following the trajectory shown in Fig.~\ref{fig:2-qubit}f. Similarly to 1-qubit $z$ gates, the electron is first displaced in a fast time scale ($\sim0.3$~ns) set by the charge qubit parameters ($\epsilon_{\rm o}$ and $V_t$), followed by a slower sweep ($\sim19$~ns) set by the spin-charge coupling parameters ($\delta_{\rm so}$ and $g_{\rm so}$), until it reaches the ionization point. The electron remains still for a short time before the whole process is then reversed. In the end a $\sqrt{i\mathrm{SWAP}}$ gate is performed. While some amount of charge is excited during the process, it goes back to its ground state, $\ket{gg}$, with an adiabatic error around $10^{-3}$.

We quantify the 2-qubit gate fidelity in presence of the most deleterious noise types for our qubits, namely quasi-static $E_z$ noise and charge-phonon relaxation. For this, we observe that the optimal gate fidelities are achieved when $E_z(\tau_{\sqrt{i\mathrm{SWAP}}}/2)\approx E_z^0$. Similarly to 1-qubit $x$-gates, this happens because $\sqrt{i\mathrm{SWAP}}$ gates are sensitive to gate time jitter, and therefore errors are minimized at the CQSS where $g_{\rm 2q}^{\rm ff}$ is robust against $E_z$ noise to first order -- recall Fig.~\ref{fig:2-qubit}f and Eq.~\ref{eq:flipdipSWAP_Delta}). An optimization algorithm finds the best adiabatic factor $K$ that minimizes errors due to $E_z$ noise for each value of $V_{t,1}=V_{t,2}=V_t$. The result is shown in Fig.~\ref{fig:2-qubit}g. Smaller detunings $\delta_{\rm so}$ (small $V_t$) result in shorter gate times, which in turn reduces errors from quasi-static noise. However, this also implies more charge states in the qubit eigenstates, which slightly increases relaxation errors. The lowest error rates, $\sim3\times10^{-3}$ are found at small detunings, $V_t-\epsilon_{\rm ff}-g_{\rm dd}\approx 100$~MHz ($V_t\approx11.59$~GHz). At smaller detunings, the 2-qubit coupling rate is too fast, which requires faster adiabatic sweeps to avoid over-rotation (lower $K$, Fig.~\ref{fig:2-qubit}g), generating more leakage errors. The gate errors remain within $10^{-3}-10^{-2}$ for a wide range of $V_t$. Finally, we estimate in Fig.~\ref{fig:2-qubit}h how noise errors depend on the noise amplitude and adiabatic factor $K$, which sets the gate time.

Our proposed 2-qubit gates are not only highly protected against noise, but also robust against donor misplacement. Indeed, variations in $r$, $d_1$ and $d_2$ mainly cause variations in $g_{\rm dd}$, therefore simply changing the energy separation between molecular charge states (Fig.~\ref{fig:2-qubit}c). This does not modify $g_{\rm 2q}^{\rm ff}$ substantially, provided that $V_t$ can be tuned accordingly (Fig.~\ref{fig:2-qubit}e) -- following \textit{e.g} the method discussed in Supplementary Information \ref{App:Vt-tunability}. A limit is reached when $D_{\rm dd}\ll\delta_{{\rm so},i}$, resulting in negligible $g_{\rm 2q}^{\rm ff}$ because of the opposite signs in Eq.~\ref{eq:flipdipSWAP_Delta}. Still, 2-qubit gates are highly effective for inter-qubit separations around $100-500$~nm (or even larger since the metallic interface on top vertical dipoles increase their interaction -- see Supplementary Information \ref{App:Screening-image-charges}) or, equivalently, two orders of magnitude tolerance in $g_{\rm dd}$. In particular, 2-qubit gate speeds do not decay with $r^3$, as opposed to standard dipole-dipole coupling schemes \cite{Ogorman2014,Hill2015}. They are similarly fast and robust for $2^{\rm nd}$- and $3^{\rm rd}$-nearest neighbors, opening up new connectivity possibilities for a large-scale quantum processor \cite{Li2017}. The large tolerance in $g_{\rm dd}$ also accommodates very well the donor depth uncertainties inherent to ion implantation \cite{Donkelaar2015}, given the linear dependence of $g_{\rm 2q}^{\rm ff}$ on $d_i$ (Eqs. \ref{eq:V_dd} and \ref{eq:g_dd}).  

We conclude that our scheme provides a dramatic reduction in the fabrication complexity, especially compared to schemes that require placing a gate \emph{between} a pair of tightly-spaced donors, such as the Kane's proposal \cite{Kane1998}, which requires $r\approx15$~nm separation between two $^{31}$P nuclear spins.

\begin{figure*}
\centering
\includegraphics[width=\textwidth]{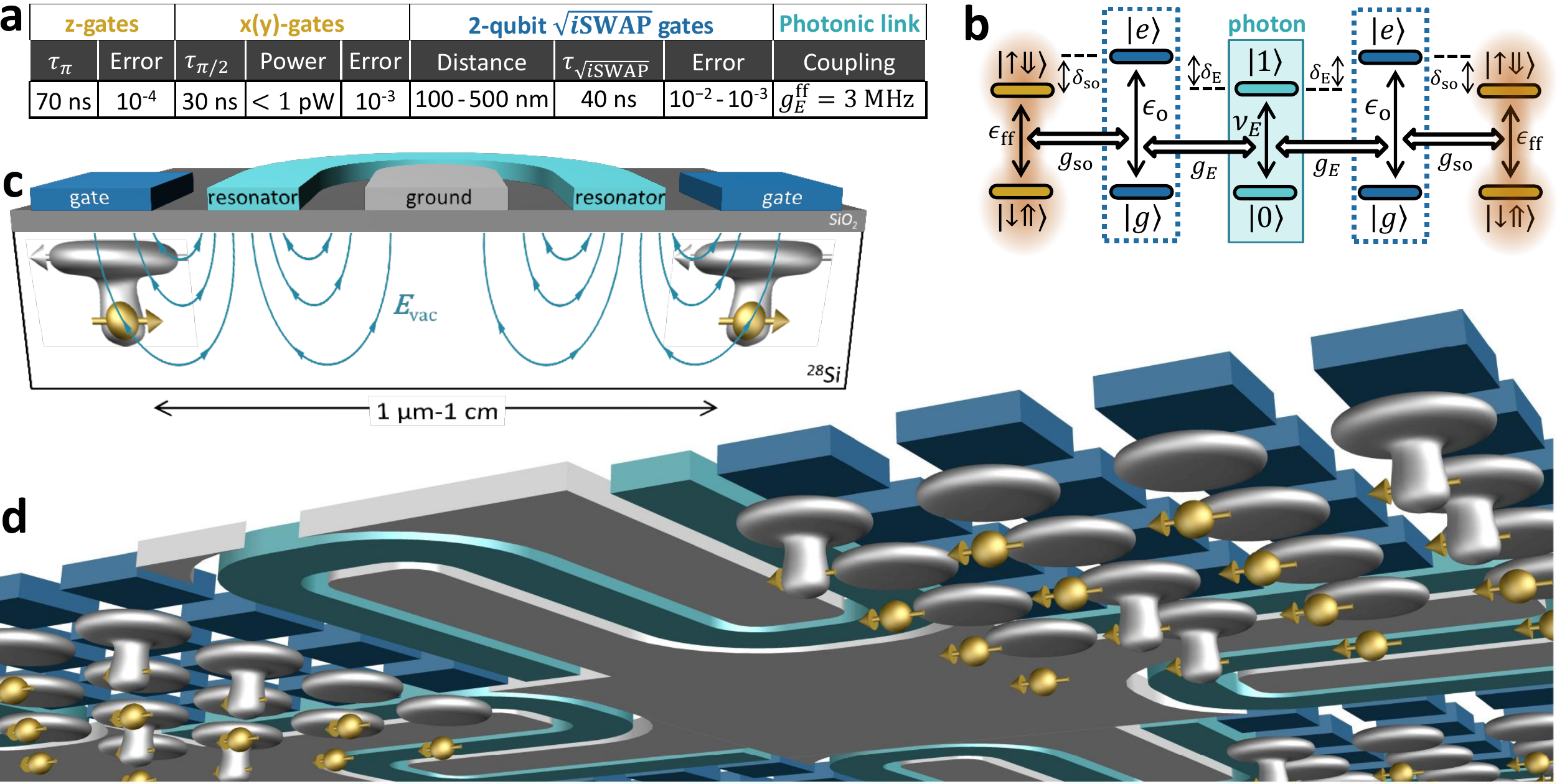}
\caption{\textbf{Silicon hybrid quantum processor}.
\textbf{a}, Figures of merit summarizing the speed and error rates of different gate schemes presented in this paper, assuming realistic noise sources. 
\textbf{b}, Level diagram for distant flip-flop qubit coupling via a microwave resonator showing first photon levels and off-resonant charge states. 
\textbf{c}, Device scheme for coupling qubits via a photonic link. Distant donors, placed next to the resonator center line and biased to their ionization point, are subject to the vacuum electric field $\textit{\textbf{E}}_{\rm vac}$ of a shared microwave resonator.
\textbf{d}, Schematic view of a large-scale quantum processor based upon $^{31}$P donors in Si, operated and coupled through the use of an induced electric dipole. Idle qubits have electron at interface, leaving the $^{31}$P nucleus in the ultra-coherent ionized state.  Electrons are partially shifted towards the donor for quantum operations. Sketch shows a possible architecture where a cluster of qubits is locally coupled via the electric dipole, and a subgroup thereof is further coupled to another cluster through interaction with a shared microwave cavity (aqua). The drawing is not to scale; control lines and readout devices are not shown.}
\label{fig:processor}
\end{figure*}

\ \\
\textbf{Scaling up using circuit quantum electrodynamics}
\vspace{1mm}

\noindent

In order to reach the long-term goal of a large-scale quantum processor, wiring up the control and read-out lines for each individual qubit is not trivial, given the high-density that spin qubits imply \cite{Vandersypen2016}. Recent solutions include cross-wiring using multilayer lithography \cite{Hill2015} or floating gate electrodes inspired by dynamic random access memory systems \cite{Veldhorst2016}. In both cases, using flip-flop qubits with long-distance interactions would result in widely spaced donors and loose fabrication tolerances. In addition, since flip-flop qubits are coupled via electric fields, they could be spaced further apart by using electrical mediators. These include floating metal gates \cite{Trifunovic2012} or even microwave resonators. Indeed, the use of electric dipole transitions allows a natural integration of donor-based spin qubits into a circuit-Quantum Electrodynamics (QED) architecture \cite{Blais2004,Childress2004,Xiang2013,Mi2016} (see Fig. \ref{fig:processor}c for a possible device layout).

A full quantum mechanical treatment yields a charge-photon coupling rate given by Eq. \ref{eq:g_E}, with $\nu_E$ now representing the resonator fundamental mode frequency and $E_{\rm ac}$ the resonator vacuum field, $E_{\rm vac}$ . Again, it is best to have the charge excited state detuned from the flip-flop transition and resonator photon (see Fig. \ref{fig:processor}b), therefore minimizing charge excitation while retaining a second-order flip-flop-photon coupling given by Eq. \ref{eq:g_E_ff}. Assuming $\delta_{\rm so}\approx\delta_E\approx 10g_{\rm so}\approx 10g_E$, a $d=15$~nm deep $^{31}$P flip-flop qubit would be coupled to photons at a $g^{\rm ff}_E\approx3$~MHz rate. This is three orders of magnitude faster than the electron-spin coupling rate to a resonator via its magnetic vacuum field \cite{Tosi2014,Haikka2017}, and comparable to the coupling strength obtained by using strong magnetic field gradients \cite{Hu2012,Viennot2015}, but without the need to integrate magnetic materials within a superconducting circuit. This assumes a vacuum field amplitude $E_{\rm vac}\approx 30$~V/m, which can be obtained by using tapered coplanar waveguide or high-inductance resonators \cite{Samkharadze2016}.

The possibility of coupling the qubits to microwave photons provides a path for dispersive qubit readout, as well as for photonic interconnects. Near-quantum limited amplifiers have recently become available to obtain excellent readout speed and fidelities \cite{Castellanos2008}. The resonator can also be used as a quantum bus to couple two spin qubits separated by as far as 1~cm (Fig. \ref{fig:processor}c), a distance given by the mode wavelength. Fig. \ref{fig:processor}b shows the detailed energy level diagram. To avoid losses from photon decay, the qubits should be detuned from the resonator by an amount much greater than the qubit-photon coupling rates. Assuming $\delta^{\rm ff}_E=10g_{E}^{\rm ff}$, where $\delta^{\rm ff}_E=\nu_E-\epsilon_{\rm ff}$, the effective 2-qubit coupling $g_{\rm 2q}^{\rm ff}\approx(g_E^{\rm ff})^2/\delta^{\rm ff}_E\approx 0.3$~MHz yields a $\sqrt{i\mathrm{SWAP}}$ gate that takes only $0.4~\mu$s.

\ \\
\textbf{Outlook: building a quantum processor}
\vspace{1mm}

\noindent
Fig. \ref{fig:processor}a summarizes the key figures of merit of a quantum processor based on flip-flop qubits coupled by electric dipole interactions. Fast 1-qubit $x$-gates are attainable with low electric drive power and error rates $\sim10^{-3}$. 2-qubit $\sqrt{i\mathrm{SWAP}}$ gates are fast and with error rates approaching $10^{-3}$. At the end of all operations, the phase of each qubit can be corrected, via adiabatic $z$-gates, in fast time scales and low error rates $\sim10^{-4}$. These values are based on current experimentally known values of charge noise in silicon devices \cite{Freeman2016}, and are possibly amenable to improvement through better control of the fabrication parameters. More advanced control pulse schemes could allow for faster gates with less leakage \cite{Motzoi2009,Ghosh2016,,Werschnik2007}, and active noise cancellation techniques, $e.g.$ pulses for gate time jitter \cite{Hill2007} or decoherence \cite{Sar2012} suppression, could further improve gate fidelities. 

Idle qubits are best decoupled from all other qubits by having the electron at the interface and the quantum state stored in the nuclear spin, which has a record coherence times $T_2 \gtrsim 30$~s (ref. \onlinecite{Muhonen2014}), and can be even longer in bulk samples \cite{Saeedi2013}. Quantum information can be swapped between the nuclear and the flip-flop qubit by simply applying an ESR $\pi$-pulse that excites the $\lvert{\downarrow\Downarrow}\rangle$ state to $\lvert{\uparrow\Downarrow}\rangle$ (Fig.~\ref{fig:A(E)}).

Qubit read-out can be obtained by spin-dependent tunneling into a cold charge reservoir, detected by a single-electron transistor \cite{Morello2010}. Read-out times can be $\sim1~\mu$s with cryogenic amplifiers \cite{Curry2015}, which is comparable to the time necessary to perform, for example, $\sim 20$ individual gates lasting $\sim 50$~ns each, in a surface code error correction protocol \cite{Fowler2012}.

A large-scale, fault-tolerant architecture can be built in a variety of ways. One- or two-dimensional arrays can be built to implement error correction schemes such as the Steane \cite{Steane1996} or the surface \cite{Fowler2012} code, since all mutual qubit couplings are tunable and gateable. A larger processor can include a hybrid of both coupling methods, incorporating cells of dipolarly-coupled qubits, interconnected by microwave photonic links (Fig.~\ref{fig:processor}), in which case more advanced error-correction codes can be implemented \cite{Knill2005,Nickerson2013,Terhal2015,Li2017}. Microwave resonators could be also used to interface donors with superconducting qubits \cite{Barends2014,Devoret2013}, for the long-term goal of a hybrid quantum processor that benefits from the many advantages of each individual architecture \cite{Xiang2013}.

In conclusion, we have presented a novel way to encode quantum information in the electron-nuclear spin states of $^{31}$P donors in silicon, and to realize fast, high-fidelity, electrically-driven universal quantum gates. Our proposal provides a credible pathway to the construction of a large-scale quantum processor where atomic-size spin qubits are integrated with silicon nanoelectronic devices, in a platform that does not require atomic-scale precision in the qubit placement. The qubits are naturally amenable to being placed on two-dimensional grids and, with realistic assumptions on noise and imperfections, are predicted to achieve error rates compatible with fault-tolerant quantum error correction.

\ \\
\textbf{METHODS} \\
\textbf{Adiabaticity}
\vspace{1mm}

Given a time-dependent Hamiltonian in a 2-dimensional Hilbert space,

\begin{equation}
\mathcal{H}_2=\Delta(t)\sigma_z+\Omega(t)\sigma_x,
\end{equation}

in units of rad/s, the adiabatic condition is expressed as \cite{Garwood2001}

\begin{equation} \label{eq:K_def}
K=\left|\frac{\omega_{\rm eff}}{\dot{\alpha}}\right|\gg1,
\end{equation}

where $\omega_{\rm eff}=\sqrt{\Delta^2+\Omega^2}$ is the instantaneous transition angular frequency between eigenstates, and $\dot{\alpha}$ is the rate of change of the orientation of $\omega_{\rm eff}$ ($\alpha=\arctan{(\Omega/\Delta)}$). It follows from Eq.~\ref{eq:K_def} that

\begin{equation} \label{eq:K_appl}
K=\frac{\left(\Delta^2+\Omega^2\right)^{3/2}}{|\dot{\Delta}\Omega-\dot{\Omega}\Delta|}\gg1,
\end{equation}

Although the processes described in this paper involve multiple levels, we applied Eq.~\ref{eq:K_appl} in different forms as an approximation of adiabaticity. This was confirmed to be always valid by checking that the leakage errors were kept below a target level.

In particular, for 1-qubit z-gates and 2-qubit $\sqrt{i{\rm SWAP}}$ gates, we used $\Delta_{\rm c}=\pi e(E_z - E_z^0)d/h$ and $\Omega_{\rm c}=\pi V_t$ to find $K_{\rm c}$ for the charge qubit, and $\Delta_{\rm so}=\pi\delta_{\rm so}$ and $\Omega_{\rm so}=2\pi g_{\rm so}$ to find $K_{\rm so}$ for the spin-charge coupling. For a chosen adiabatic factor $K$, we find $E_z(t)$ by satisfying the condition $\min(K_{\rm so},K_{\rm c})=K$.

For 1-qubit drive, we used $\Delta_E=\pi\delta_E$ and $\Omega_{E}=2\pi g_E$ to find $K_E$. A particular choice of $K=K_E$ sets the adiabatic sweep rate of $E_{\rm ac}(t)$.

\ \\
\textbf{Estimation of dephasing and gate errors}
\vspace{1mm}

\noindent
In order to estimate the effects of quasi-static $E_z$ noise on dephasing, we first calculate the flip-flop qubit transition frequency $\epsilon_{\rm ff}$ (difference between eigenfrequencies corresponding to eigenstates closest to $\ket{g\downarrow\Uparrow}$ and $\ket{g\uparrow\Downarrow}$, which we denote as $\ket{g\downarrow\Uparrow}_{\rm e}$ and $\ket{g\uparrow\Downarrow}_{\rm e}$). Next, for an equally distributed noise range $E_z^n=\sqrt{3}[-E_{z,\rm rms}^{\rm noise},E_{z,\rm rms}^{\rm noise}]$, we estimate the qubit dephasing rate to be 

\begin{equation}
{\rm Dephasing~rate}=\sum\limits_{n}{\left|\epsilon_{\rm ff}-\epsilon_{\rm ff}^n\right|/N_n},
\end{equation}

where $N_n$ is the number of sampled $E_z^n$ and $\epsilon_{\rm ff}^n$ is calculated for each value of $E_z^n$. 

The averaged error rate (without noise) of a desired adiabatic unitary process $U_{\rm ideal}$ is calculated by averaging the fidelity of the actual process $U$ over a set of initial states $\ket{j}$,

\begin{equation}
{\rm Adiabatic~error}=1-\sum\limits_{\ket{j}}{\left|\bra{j}U^\dagger U_{\rm ideal}\ket{j}\right|^2/N_j},
\end{equation}

where $N_j$ is the number of initial states. For 1-qubit gates ($e.g.$ a $\pi$ z-gate or a $\pi/2$ x(y)-gate), we choose $\ket{j}=\{\ket{g\downarrow\Uparrow}_e,\ket{g\uparrow\Downarrow}_e,(\ket{g\downarrow\Uparrow}_e+\ket{g\uparrow\Downarrow}_e)/\sqrt{2},(\ket{g\downarrow\Uparrow}_e+i\ket{g\uparrow\Downarrow}_e)/\sqrt{2}\}$ and $N_j=4$, whereas for 2-qubit gates ($e.g.$ $\sqrt{i{\rm SWAP}}$) $\ket{j}=\ket{j}_1\otimes\ket{j}_2$ (the ${1,2}$ indexes refer to the aforementioned 4 initial states for each qubit) and $N_j=16$.

To estimate the averaged gate error rate under quasi-static $E_z$ noise, the actual process $U$ and eigenstates $\ket{j}$ are calculated for each value of $E_z^n$ before averaging,

\begin{equation}
{\rm Noise~error}=1-\sum\limits_{n,\ket{j}_n}{\left|\bra{j}_nU_n^\dagger U_{n,\rm ideal}\ket{j}_n\right|^2/(N_jN_n)}
\end{equation}

Finally, to estimate errors due to charge-phonon relaxation, we multiply the averaged charge excitation by its relaxation rate and assume a exponential decay in fidelity:

\begin{equation}
{\rm Relax.~error}=\frac{1-e^{-\int\limits_{0}^{\tau_{\rm gate}}\left({\sum\limits_{\ket{j(t)}}{\braket{j(t)|e}\braket{e|j(t)}/N_j
}}\right)dt/T_{1, \rm o}}}{2},
\end{equation}

where $\ket{j(t)}$ are the time-evolution of the initial set states $\ket{j}$. For 2-qubit gates, we sum up the error rate of each qubit.

\vspace{3mm}
\textbf{Acknowledgments} We thank A. Blais, H. Bluhm, M. Eriksson, J. O'Gorman, S. Benjamin, J. Salfi, M. Veldhorst, A. Laucht, R. Kalra and C. A. Parra-Murillo for discussions. This research was funded by the Australian Research Council Centre of Excellence for Quantum Computation and Communication Technology (project number CE110001027), the US Army Research Office (W911NF-13-1-0024) and the Commonwealth Bank of Australia. Tight-biding simulations used NCN/nanohub.org computational resources funded by the US National Science Foundation under contract number EEC-1227110.
\newline \\
\textbf{Author Contributions}
A.M. and G.T. conceived the project.
G.T. developed the theoretical framework, with F.A.M.'s assistance and under A.M.'s supervision.
G.T. and F.A.M. performed calculations and numerical simulations with S.T.'s and V.S.'s assistance.
R.R. and G.K. developed the qubit simulation capabilities in the NEMO-3D code.
G.T., A.M. and F.A.M. wrote the manuscript, with input from all co-authors.
\newline \\
\textbf{Additional Information} Supplementary information accompanies the paper. Correspondence and requests for materials should be addressed to G.T. (g.tosi@unsw.edu.au) or A.M. (a.morello@unsw.edu.au).

\clearpage

\onecolumngrid
\begin{center}
\textbf{SUPPLEMENTARY INFORMATION for \\
``Silicon quantum processor with robust long-distance qubit couplings", by G. Tosi \emph{et.al.}\\
\vspace{3mm}}
\end{center}
\twocolumngrid

\renewcommand\thesubsection{S\arabic{subsection}}
\setcounter{page}{1}
\renewcommand*{\thepage}{S\arabic{page}}
\setcounter{equation}{0}
\renewcommand{\theequation}{S\arabic{equation}}
\setcounter{figure}{0}
\renewcommand{\thefigure}{S\arabic{figure}}
\renewcommand*{\citenumfont}[1]{S#1}
\renewcommand*{\bibnumfmt}[1]{[S#1]}

\subsection{\label{App:Nemo-orb}Validity of the two-level approximation for the electron orbital wavefunction}

The concepts and calculations shown in the manuscript are based upon approximating the electron orbital degree of freedom as a two-level system, i.e. a charge qubit. The true orbital levels of a donor-interface system are, of course, more complex than that. However, below we show that the charge qubit model represents an excellent approximation, for the range of parameters relevant to our proposal.

\begin{figure}
\centering
\includegraphics[width=\columnwidth]{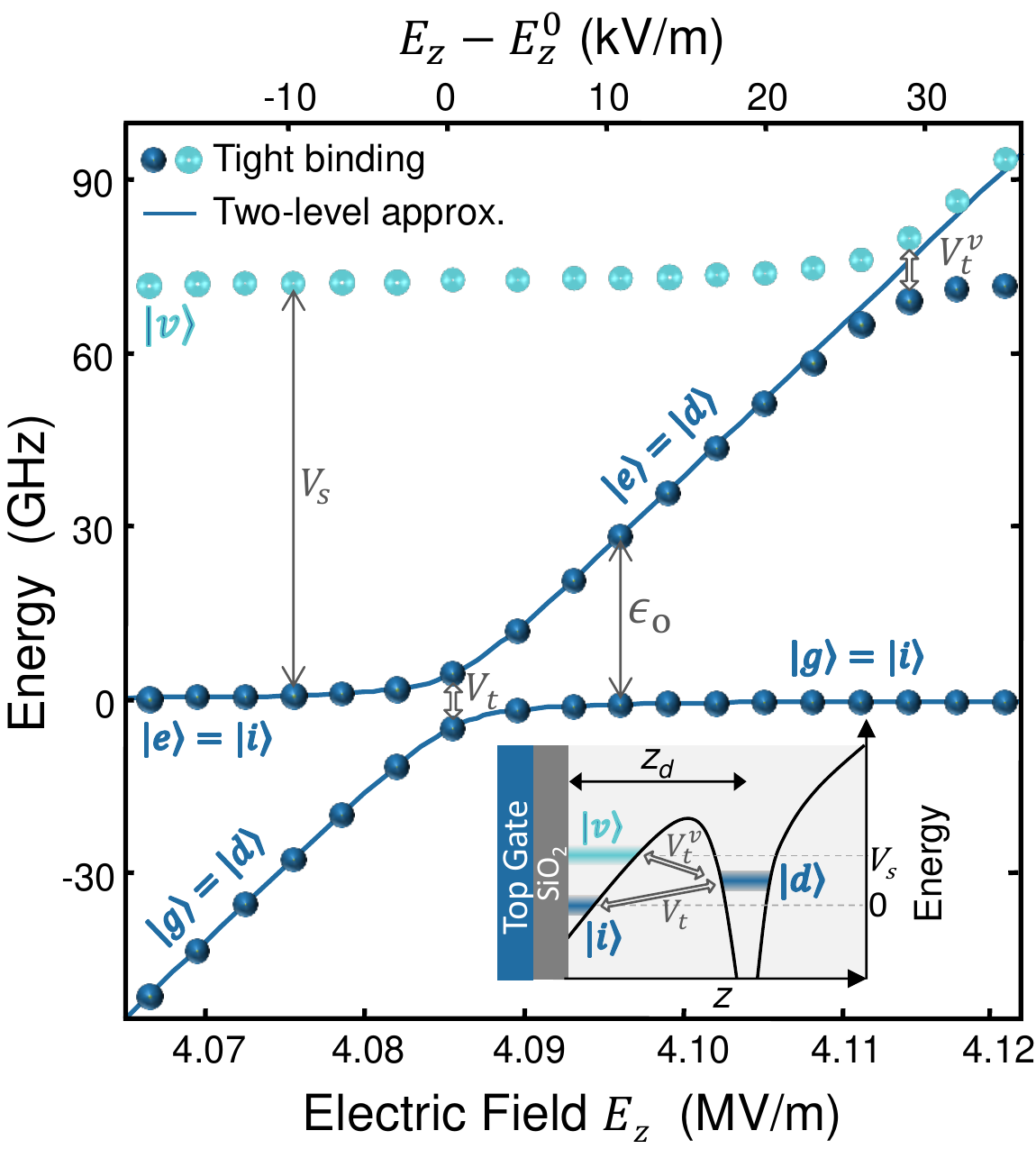}
\caption{\textbf{Orbital and valley states.} The lowest orbital energy levels of the donor-interface system, with respect to the lower valley interface state $\lvert i \rangle$ (set as the zero-energy reference). The donor is assumed 15.2 nm below a Si/SiO$_2$ interface. The dots correspond to the energy levels obtained from a full-scale tight-binding calculation with NEMO-3D. Solid lines represent the energy levels obtained from the two level approximation described by Eq. \ref{eq:H_orb} in the main manuscript. An excellent agreement between our two-level model and tight binding calculations is observed, since the valley splitting $V_s$ is much larger than the tunnel coupling $V_t$. Inset: Potential profile as a function of depth, illustrating the donor $\lvert d \rangle$, lower $\lvert i \rangle$ and upper $\lvert v \rangle$ valley interface states. The donor ground state is tunnel-coupled to the lower and upper valley interface states by $V_t$ and $V_t^v$  respectively.}
\label{fig:Nemo-levels}
\end{figure}

The ground orbital wavefunction $\lvert d \rangle$ of an electron bound to a donor is a symmetric combination of the 6 conduction band minima (``valleys'') ($\mathrm{k_{\pm x}}$, $\mathrm{k_{\pm y}} $, $\mathrm{k_{\pm z}} $) in silicon\cite{Kohn1955S}. Higher excited valley-orbit states are separated by $>10$~meV and can be safely neglected. Conversely, the orbital states of an electron confined at the Si/SiO$_2$ interface comprise a low-energy doublet of states, with wavefunctions constructed as a combination of the $\mathrm{k_{\pm z}}$ valleys. The $\mathrm{k_{+z}}$ and $\mathrm{k_{-z}}$ valleys are coupled by the abrupt potential of the interface, which breaks the degeneracy of the ground state doublet into the lower valley $\lvert i \rangle$ and upper valley $\lvert v \rangle$  states, separated by the valley splitting $V_s$ \cite{Saraiva2009S}. All the remaining excited donor and interface orbital states are well above the ground doublet by several meV \cite{Rahman2009S, Calderon2009S}. When the donor is close to ionization, the lowest-energy states of the system therefore consist of $\lvert d \rangle$, $\lvert i \rangle$ and $\lvert v\rangle$ states, as shown in Fig.~\ref{fig:Nemo-levels} inset.

We computed the above three energy levels with the atomistic tight binding package NEMO-3D\cite{Klimeck2007S, Klimeck2007aS}, assuming a donor placed at depth $z_{\rm d} = 15.2$~nm below the Si/SiO$_2$ interface, and biased close to the donor ionization field $E_z^0$. The dependence of the energies of $\lvert d \rangle$, $\lvert i \rangle$ and $\lvert v\rangle$ on electric field $E_z$ is shown by the dots in Fig. \ref{fig:Nemo-levels}. We also fit the lowest energy levels with the charge qubit two-level model described by the Hamiltonian $\mathcal{H}_{\rm orb}$ (in Eq. \ref{eq:H_orb} of the main manuscript), and plot them as solid blue lines in Fig. \ref{fig:Nemo-levels}. The two-level model agrees well with tight-binding calculation taking $V_t = 9.3$~GHz and $d = 11$~nm in Eq.~\ref{eq:H_orb}. Here, $d$ represents the separation between the center-of-mass positions of the donor-bound ($\ket{d}$) and interface-bound ($\ket{i}$) orbitals. This is the relevant quantity in calculating the electric dipole strength. The extracted value $d$ is lower than the donor depth $z_{\rm d}$, as expected, and is consistent with the separation between the mean positions of the donor and interface electron wavefunctions as modeled with NEMO-3D.

Fig.~\ref{fig:Nemo-levels} shows that, when $E_z \ll E_z^0$, the orbital ground state $\lvert g \rangle $ of the electron is localized at the donor, whereas the first excited state corresponds to the lower valley interface state. The two states are separated in energy by $\epsilon_o$, given by Eq.~\ref{eq:e_o} of the main manuscript. As $E_z$ increases, the two states approach, and anticross at $E_z = E_z^0$. For $E_z \gg E_z^0$, the donor state will eventually (at $E_z^v \sim 4.11$~MV/m) anticross with the upper valley interface state. Therefore, as shown by the solid lines in Fig.~\ref{fig:Nemo-levels}, a two-level model described by the $\lvert d \rangle$ and $\lvert i \rangle$ states constitutes an excellent approximation for $E_z < E_z^v$. This allows a broad range of validity of the simple charge qubit model, provided the interface valley splitting $V_s$ is much larger than the tunnel coupling $V_t$. The NEMO-3D model used here predicts $V_s = 71.7$~GHz, which is indeed much larger than $V_t = 9.3$~GHz. Experimentally, even higher values of $V_s$ are routinely observed in electrons confined at the Si/SiO$_2$ interface by top-gated structures \cite{Yang2013S}, providing further reassurance on the practical validity of our models.

\subsection{\label{App:Vt-tunability}Tunnel coupling tunability between donor and interface}

The tunnel coupling $V_t$ of the electron between the donor and interface orbital states plays a key role in our models. It influences all the driving strengths (Eqs.~\ref{eq:g_so}, \ref{eq:g_E_ff} and \ref{eq:g_E}) and inter-qubit couplings (Eq.~\ref{eq:flipdipSWAP_Delta}). In the presence of a single metal gate above the donor location, the dependence of $V_t$ on donor depth has been analyzed with effective mass theory \cite{Calderon2008S,Calderon2009S}. Ref.~\onlinecite{Calderon2008S} indicates that $V_t$ depends exponentially on donor depth $z_d$, and decreases by an order of magnitude for every 6~nm increase in $z_d$. Moreover, in addition to the exponential decay, the tunnel coupling also has an oscillatory dependence on $z_d$ at the atomic scale due to valley interference effects \cite{Calderon2008S}.

Using the ion implantation technique, the placement of a donor at $z_d \approx 15$~nm below the interface with a 5~nm thick oxide results in a vertical uncertainty of order $\pm 10$~nm (ref.~\onlinecite{Donkelaar2015S}), resulting in more than 2 orders of magnitude uncertainty in $V_t$. Therefore, it is crucial to implement a method to tune $V_t$ \emph{in situ}.

\begin{figure}
\centering
\includegraphics[width=\columnwidth]{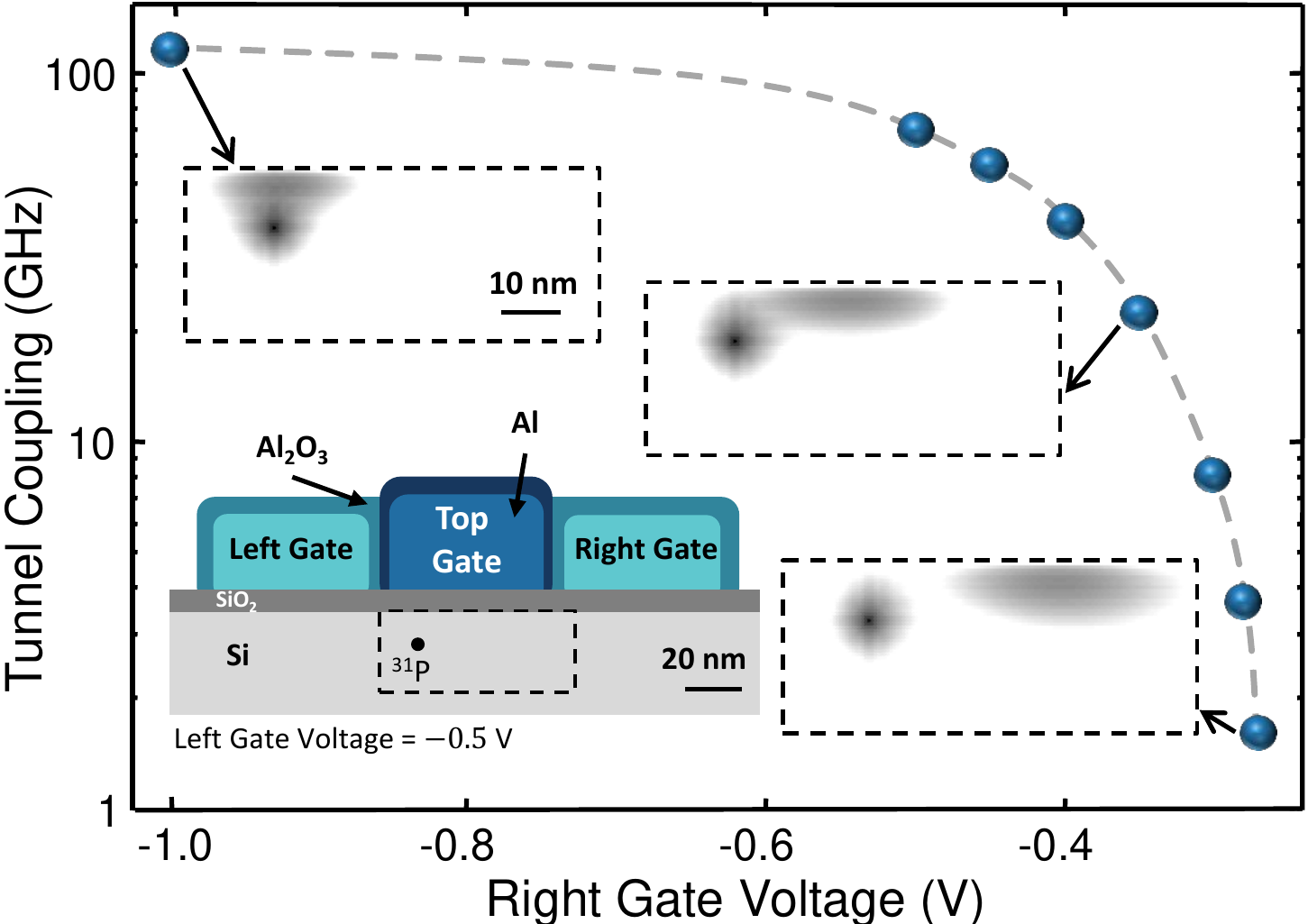}
\caption{\textbf{Gate-tunability of the tunnel coupling.} Tunnel coupling $V_t$ of the charge qubit as a function of gate voltage. To tune $V_t$, additional gates (left and right) are present on either side of the top gate which pulls the $\mathrm{^{31}P}$ donor electron to the interface. The insets illustrate the NEMO-3D wavefunctions, when $V_r =$ -1, -0.35 and -0.27 V. $V_l = -0.5$ V for all the simulations, and the top gate is biased such that the position of the electron is in between the donor and interface. The donor is assumed to be $z_d=9.2$~nm below a SiO$_2$ interface.}
\label{fig:tunnel_coupling_tunability}
\end{figure}

Here, we propose that $V_t$ can be controlled by adding two gates (left and right) on either side of the gate (top) which pulls the donor electron to the interface. The relative voltages $V_l$ and $V_r$ applied to the left and right gates respectively can modify the potential landscape, and displace laterally the location of the interface wavefunction. This, in turn modifies the distance between the donor and interface wavefunctions allowing $V_t$ to be significantly reduced. We use a combination of a finite element Poisson solver \cite{Note1} and NEMO-3D to estimate $V_t$ in this device topology. In Fig.~ \ref{fig:tunnel_coupling_tunability} we plot the tunnel coupling tunability as a function of $V_r$, assuming $V_l = -0.5$~V and $z_d = 9.2$~nm.  The insets of Fig.~ \ref{fig:tunnel_coupling_tunability} show the NEMO-3D electron wavefunctions, when the top gate is biased such that the mean position of the electron is in between the donor and interface. We infer that the electron wavefunction at the interface can be moved by several tens of nanometers with $V_r$, allowing $V_t$ to be tuned by at least $\sim 2$ orders of magnitude. This technique therefore enables us to circumvent the uncertainty in donor depth and $V_t$ arising from ion-implantation, while remaining straightforward from a nanofabrication point of view. Note that, by relocating the problem of valley oscillations from the exchange interaction (Kane proposal) to the tunnel coupling (our proposal), we have effectively provided a way in which the delicate parameter can now be tuned using a much simpler gate geometry. Indeed, the gate layout used in this model is essentially identical to the layout routinely adopted for the fabrication of electrostatically-defined quantum dots at the Si/SiO$_2$ interface \cite{Yang2013S,Veldhorst2014S}. These results indicate that a viable strategy for the construction of an ion-implanted quantum processor based upon our idea is to aim for an implantation depth that is by default rather shallow, then reduce $V_t$ locally with the use of the surface gate stack.

\subsection{\label{App:Elec_noise}Charge and gate noise}

In the main manuscript, we have presented estimates of dephasing rates and gate errors extracted from models where we assume a quasi-static (i.e. with a spectral weight centered at frequencies smaller than the qubit resonance and the Rabi frequency) electric field noise acting on the qubits. Here we explain why this assumption, and the r.m.s. value of 100 V/m for the noise, is justified for silicon nanoelectronic devices.

Given that the distance between the donor and interface sites is $\sim10$-30~nm, a vertical noise field of 100~V/m would correspond to 1-3~$\mu$eV charge detuning noise. This is consistent with the 1-9~$\mu$eV noise found in a range of semiconductor nanodevices, including SiGe \cite{Kim2015S,Thorgrimsson2016S,Freeman2016S}, AlGaAs \cite{Dial2013S} and Si/SiO$_2$ \cite{Harvey-Collard2015S,Freeman2016S}. In particular, MOS structures where found recently to have similar charge noise levels as SiGe devices, around 1.5~$\mu$eV.

The particular geometry of our qubits contributes to making them less susceptible to device-intrinsic charge noise. First of all, the electric dipole induced on a donor 15 nm below the Si/SiO$_2$ interface is substantially smaller than that of lateral gate-defined double quantum dots. Second, our qubits are largely insensitive to horizontal charge noise. Indeed, the orientation of the donor-interface dipole is mostly vertical (even when the interface wavefunction is displaced laterally, since image charges screen the lateral dipole -- see Supplementary Information \ref{App:Screening-image-charges}). The only effect of lateral noise is to modulate $V_t$ -- see Supplementary Information \ref{App:Vt-tunability}. For $V_t\approx10$~GHz, Fig. \ref{fig:tunnel_coupling_tunability} suggests that $10~\mu$V r.m.s. lateral noise would cause less than 1\% uncertainty in $\delta_{\rm so}$ (and therefore in gate time), which translates into maximum $10^{-4}$ errors due to gate time jitter for the flip-flop qubit, well below other contributions,  and maximum $\sim10^4$~Hz extra dephasing due to dispersive shifts (Eq. \ref{eq:Dorb}).

Another source of vertical electric field noise can be the thermal and electrical noise produced by the metallic gates on top of the qubits, and the room-temperature instruments they connect to. An $R=50~\Omega$ resistor at room temperature produces Johnson-Nyquist noise with an r.m.s voltage $\sqrt{4k_BTR\Delta\nu}$. Therefore a quasi-static bandwidth $\Delta\nu\sim10^6$~Hz produces $\sim1~\mu$V voltage noise, which is equivalent to $E_{z,\rm rms}^{\rm noise}\sim10$~V/m, or errors $<10^{-5}$ (Fig.~\ref{fig:1-qubit}g). Furthermore, because of the very low powers required by the electrically-driven 1-qubit gates and adiabatic shuttling, it is possible to insert abundant low-temperature attenuation along the high-frequency lines, and therefore the relevant temperature for the Johnson-Nyquist noise is well below room temperature. On the other hand, being close to a metallic interface, our qubit will be subject to evanescent wave Johnson noise (EWJN) due to vacuum fluctuations. Assuming the qubit is $z=15$~nm under aluminum gates at $T=100$~mK ($\sigma=1.4\times10^8$~S/m conductivity \cite{Dehollain2013S}), a quasi-static bandwidth $\Delta\nu\approx10^6$~Hz produces \cite{Henkel1999S} $\sqrt{k_BT\Delta\nu/(2z^3\sigma)}\sim0.04$~V/m r.m.s. electric field noise, therefore negligible. We conclude that the main source of quasi-static noise will be charge noise with a typical $1/\nu$ spectrum. To get $E_{z,\rm rms}^{\rm noise}=100$~V/m over a $10^6$~Hz bandwidth, the power spectral density has to be $S_{\rm c}(\omega)\approx10^4/(6\omega)$, in units of ${\rm (V/m)^2/(rad.s^{-1})}$.

So far we have only considered quasi-static noise. The presence of some residual amount of high-frequency noise could possibly lead to errors while performing quantum operations. Below we discuss these high-frequency sources, finding that they will cause much smaller errors compared to quasi-static noise.

In general, a driven qubit Rabi-oscillates with a decay envelope function given by \cite{Bylander2011S} $\zeta(t)\exp(-\Gamma_Rt)$, where $\zeta(t)$ represents decay due to quasi-static detuning noise and $\Gamma_R$ the exponential Rabi decay rate, which combines the qubit relaxation rate, $\Gamma_1$, the inverse of the gate time jitter due to quasi-static noise, $\Gamma_1^{\Delta}$, the inverse of the gate time jitter due to noise at the drive frequency, $\Gamma_1^{\nu}$, (the last three yield $T_{2\rho}$ in the dressed qubit picture \cite{Laucht2016S}) and the decay rate due to detuning noise at the Rabi frequency, $\Gamma_\Omega$ (which equals the inverse of $T_{1\rho}$ in the dressed qubit picture \cite{Yan2013S,Laucht2016S}).

The effects of $\zeta(t)$, $\Gamma_1$ and $\Gamma_1^{\Delta}$ have already been discussed extensively in this manuscript, with corresponding error levels below $10^{-3}$. We now focus on errors due to high-frequency noise sources, corresponding to decay rates $\Gamma_1^{\nu}$ and $\Gamma_\Omega$.

Vertical (thus parallel to the driving field $E_{\rm ac}$) noise at the qubit resonance frequency ($\sim 10^{10}$~Hz) would cause transitions between the qubit eigenstates -- essentially a spurious excitation/relaxation process driven by noise -- at a rate $\Gamma_1^{\nu}$. During gate operations, the portion of the noise spectrum 
around the qubit frequency adds incoherently to the external resonant drive, and causes the gate time to fluctuate. This noise can be caused $e.g.$ by vertical dipoles fluctuating in resonance with the qubit or by voltage noise at the metallic gates. For the flip-flop qubit, the Rabi decay rate is given by $\Gamma_1^\nu=(\pi/2)(\mu_e^{\rm ff}/\hbar)^2S(2\pi\epsilon_{\rm ff})$, where $\mu_e^{\rm ff}=ed\langle g_{\rm so}/\delta_{\rm so}\rangle$ is the average flip-flop qubit electric dipole moment and $S(2\pi\epsilon_{\rm ff})$ is the noise power spectral density at the qubit angular frequency (in units of ${\rm (V/m)^2/(rad.s^{-1})}$). In case of charge noise, $S_{\rm c}(\omega)=10^4/(6\omega)$, which gives $\Gamma_1^\nu\sim10^4$~Hz. This implies $\pi/2$ $x$-gate errors $\sim10^{-4}$. In case of Johnson-Nyquist noise, $S_{\rm JN}(\omega)= 2\times10^{14}R\hbar\omega\pi^{-1}(e^{\hbar\omega/k_BT}-1)^{-1}$ (where we have used $\partial E_z/\partial V=10^7~{\rm m^{-1}}$, typical in MOS nanostructures). Because of the very low powers required by the electrically-driven 1-qubit gates ($<1$~pW), it is possible to insert abundant low-temperature attenuation along the high-frequency lines, insuring that the gates are well thermalized, and the noise of the room-temperature electronics greatly attenuated. A noise temperature $T=100$~mK would give $\Gamma_1^\nu<10^4$~Hz, and therefore error rates $<10^{-4}$. Finally, in case of EWJN at $T=100$~mK, the $10^{10}$~Hz part of the spectrum is \cite{Henkel1999S,Poudel2013S} $S_{\rm EW}(\omega)\approx\hbar\omega/(4\pi z^3\sigma)$. This would give $\Gamma_1^\nu<10^4$~Hz, therefore again error rates $<10^{-4}$.

Noise at the Rabi frequency ($\Omega_R>10^{7}$~Hz) causes decay in the Rabi oscillations at a rate $\Gamma_\Omega$. This type of noise feeds into the driven qubit via fluctuations in the detuning between drive frequency and the qubit precession frequency. The decay rate of the flip-flop qubit is given by $\Gamma_\Omega=(\pi/2)(2\pi\sum_{i=x,y,z}\partial\epsilon_{\rm ff}/\partial E_i)^2S(\Omega_R)$. At the low-error operation region of Fig. \ref{fig:1-qubit}f, $\partial\epsilon_{\rm ff}/\partial E_z\sim10^3~{\rm HzV^{-1}m}$ and $\partial\epsilon_{\rm ff}/\partial E_{x,y}\sim10^2~{\rm HzV^{-1}m}$ (from Fig. \ref{fig:tunnel_coupling_tunability}). $1/\nu$ charge noise gives $\Gamma_\Omega<10^4$~Hz, implying $<10^{-4}$ errors. Johnson-Nyquist noise from room temperature gives $\Gamma_\Omega=3\times10^2$~Hz, whereas EWJN at 100~mK gives $\Gamma_\Omega=2\times10^1$~Hz, therefore producing $<10^{-5}$ and $<10^{-6}$ errors, respectively.

We conclude that the sources of error treated in the main text, namely quasi-static $E_z$ noise and charge-phonon relaxation, are the most deleterious ones for flip-flop qubits. Therefore our analysis is sufficient to provide a reliable estimate of dephasing and gate errors. Indeed, low-frequency noise was found to be the most deleterious one in a hybrid donor-dot qubit in a silicon MOS device \cite{Harvey-Collard2015S}. Finally, note that we do not assume any type of dynamical noise correction or cancellation to be applied, and therefore our calculations are a worst-case scenario.

\subsection{\label{App:Screening-image-charges}Screening effect of metals and dielectrics}

Our device topology consists of a SiO$_2$ layer sandwiched between a metal gate and silicon substrate, with the donor embedded in the substrate. In such a topology, the image charges of the donor electron and nucleus will be located above the donor, thereby creating an additional vertical dipole. In this section, we quantify the variation of the dipolar coupling $g_{\rm dd}$ due to the electric field from the additional dipole, and arrive to the conclusion that $g_{\rm dd}$ will most likely be enhanced.

\begin{figure}
\centering
\includegraphics[width=\columnwidth]{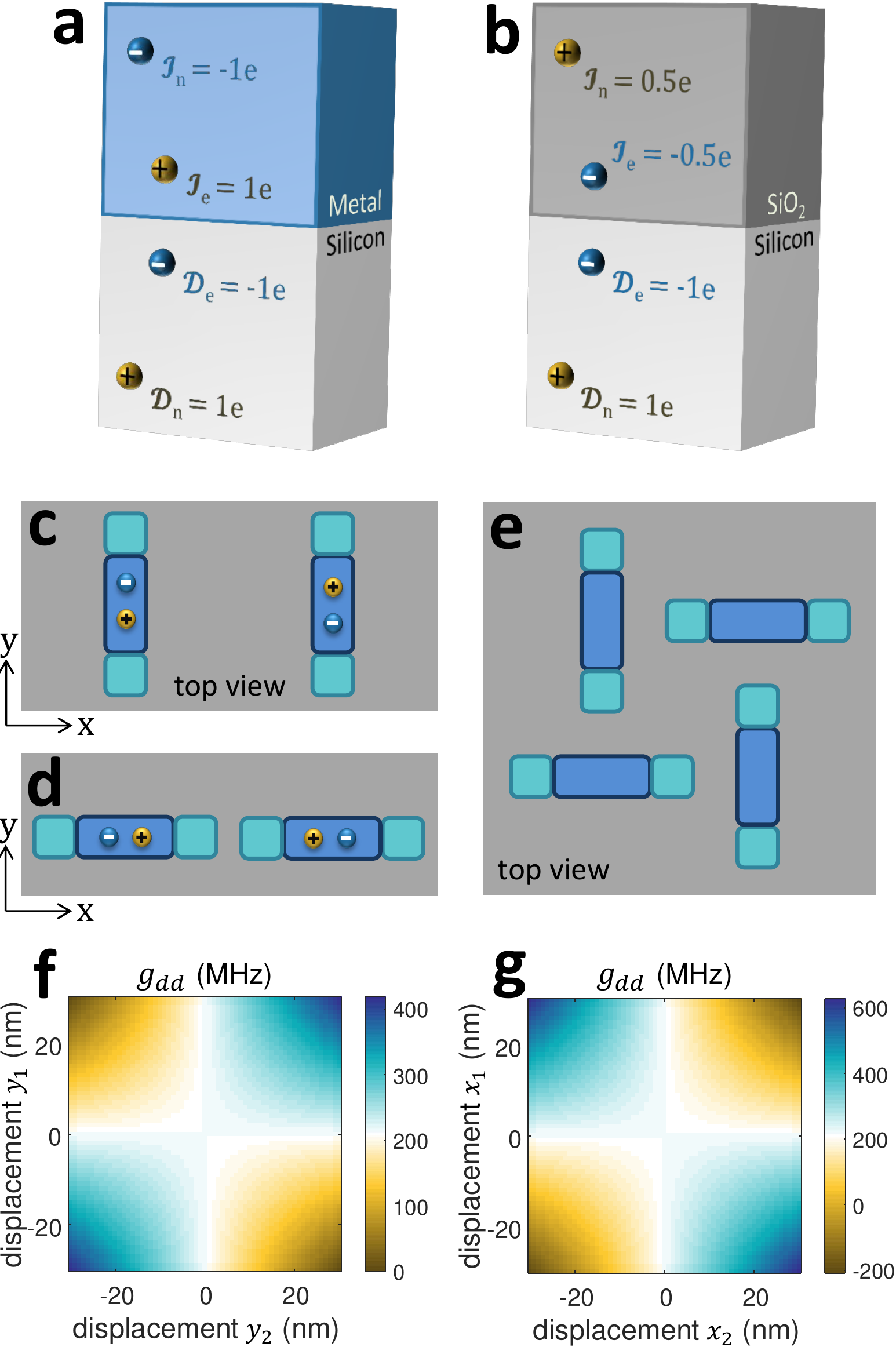}
\caption{\textbf{Screening and image charges.} Image ($\mathcal{I}_\mathrm{e}$ and $\mathcal{I}_\mathrm{n}$) charges of the donor electron ($\mathcal{D}_\mathrm{e}$) and nucleus ($\mathcal{D}_\mathrm{n}$) for silicon-metal (\textbf{a}) and silicon-oxide (\textbf{b}) interfaces. The magnitude and polarity of the image charges are given by Eq. \ref{eq:ImageCharge}. Schematic top view of two interacting dipoles when the negative charges (blue spheres) are displaced in perpendicular (\textbf{c}) and parallel (\textbf{d}) direction to the inter-dipole separation. \textbf{e}, Top view of gate stack that tunes each qubit's $V_t$ by displacing their interface states perpendicularly to their nearest neighbor displacement, leaving $g_{\rm dd}$ unchanged. Inter-dipole coupling $g_{\rm dd}$, as predicted by Eq. \ref{eq:g_dd_ic}, for the orientation shown in \textbf{c} (\textbf{f}) and \textbf{d} (\textbf{g}), for $r=200$~nm, $d_1=d_2=10$~nm and $Q=-0.5$.}
\label{fig:image_charge}
\end{figure}

The magnitude and polarity of the image charges depend on the details of the nanostructure, such as the donor depth and thickness of the oxide. We first analyze two extreme scenarios considering image charges at (i) silicon-metal and (ii) silicon-oxide interfaces. For a source donor electron (or nuclear) charge $\mathcal{D}_\mathrm{e(n)}$, in silicon, the image charge $\mathcal{I}_\mathrm{e(n)}$ in the interface material is given by\cite{Rahman2009S}

\begin{subequations}
\begin{equation} \label{eq:ImageCharge}
\mathcal{I}_\mathrm{e(n)}=Q~\mathcal{D}_\mathrm{e(n)},
\end{equation}
\begin{equation} \label{eq:Q}
Q=\frac{\epsilon_{\rm Si} - \epsilon_{\rm I}}{\epsilon_{\rm Si} + \epsilon_{\rm I}},
\end{equation}
\end{subequations}

where $\epsilon_{\rm Si} =$ 11.7 is the dielectric constant of silicon,  $\epsilon_{\rm I} =$ 3.9 and  $\infty$  for oxide and metal interfaces respectively. Figures \ref{fig:image_charge}a,b show the magnitude and polarity of the image charges for both types of interfaces. For simplicity, we assume in Fig. \ref{fig:image_charge} and Eq. \ref{eq:ImageCharge} that the donor electron as well as its image are point charges. Given that the separation between the two donors is at least 180 nm (more than hundred times the Bohr radius of the donor electron), the above assumption is valid when calculating their dipolar interaction. 

We first consider the electric dipole to be vertical. For the silicon-metal interface in Fig. \ref{fig:image_charge}a, $Q=-1$ and therefore the image charges have the opposite sign and same magnitude as the source charges. As a result, the total electric field  $E_{\rm dip}$ from each donor will be enhanced by a factor of 2. This improves the electric dipole coupling $g_{\rm dd}$ between the two donors by a factor of 4. On the contrary, for the silicon-oxide interface in Fig. \ref{fig:image_charge}b, the image charges have the same sign and reduced magnitude ($Q=0.5$) as the source charges, which decreases $E_{\rm dip}$ by half and therefore $g_{\rm dd}$ to a quarter of its bare value.

For a real device, which typically contains a few metal gates on top of a $\sim8$~nm thick SiO$_2$, it is difficult to make a precise estimate of the extra electric field from image charges. Rahman \textit{et. al.} \cite{Rahman2009S} assumed that a combination of metallic and oxide screening effects yields $Q=-0.5$, corresponding to an improvement in the magnitude of the electric dipole by $\approx$ 50\%, which yields an improvement in $g_{\rm dd}$ by 125\%. 
This means that, while building a real device, one would have to aim for slightly larger inter-donor separations than the ones presented in the main text.

Since the donor-interface tunnel coupling $V_t$ has to be tuned to a precise value, the dipole will also have lateral components as shown on the insets of Fig. \ref{fig:tunnel_coupling_tunability}. These components will also be affected by image charges. In the case of a metallic interface, Fig. \ref{fig:image_charge}a, the lateral image dipole has opposite direction as the original one, and therefore the total lateral component will be completely screened. On the other hand, for the SiO$_2$ interface, Fig. \ref{fig:image_charge}b, the lateral component will be enhanced by 50\%. Finally, for our assumed real structure ($Q=-0.5$), the lateral dipole will decrease to half its original value.

In more detail, the electric field of a donor-interface state will be the one produced by a dipole that includes both screening and angular effects,

\begin{equation} \label{eq:D_ic}
\textbf{\textit{D}}_i=\textbf{\textit{d}}_i+Q\times(d_{i,x},d_{i,y},-d_{i,z}),
\end{equation}

where $\textbf{\textit{d}}_i$ refers to the bare dipole, with $x$, $y$ and $z$ components $d_{i,x}$, $d_{i,y}$ and $d_{i,z}$, respectively. We then modify the dipole-dipole interaction term, Eq. \ref{eq:g_dd}, to \cite{Ravets2014S}:

\begin{equation} \label{eq:g_dd_ic}
g_{\rm dd}=\frac{e^2}{16\pi\varepsilon_0\varepsilon_r h}\frac{\textbf{\textit{D}}_1\cdot\textbf{\textit{D}}_2-3(\textbf{\textit{D}}_1\cdot\textbf{\textit{r}})(\textbf{\textit{D}}_2\cdot\textbf{\textit{r}})/r^2}{r^3},
\end{equation}

which includes image charges and angular dependencies. Note that we neglect the interaction of a dipole with its own charge since it does not produce inter-donor coupling.

Laterally displacing the interface charge is, in general, necessary for the purpose of tuning the donor-interface tunnel coupling $V_t$. The same displacement, however, also alters the total electric dipole and can therefore affect the dipole-dipole coupling $g_{dd}$ between neighboring qubits. We first consider the case in which the displacements are perpendicular to the separation between dipoles, Fig. \ref{fig:image_charge}c. The $g_{\rm dd}$ dependence on $y_1$ and $y_2$ is plotted in Fig. \ref{fig:image_charge}f, for maximum displacements of 30~nm (enough to tune $V_t$ by two orders of magnitude -- see Fig \ref{fig:tunnel_coupling_tunability}). It shows that, provided that the interface states are displaced along the same direction, $g_{\rm dd}$ only varies by a factor of two. For completeness, we also analyze the case in which the interface states are displaced in the same direction as the inter-donor separation (Fig. \ref{fig:image_charge}d). As can be seen in the plot in Fig. \ref{fig:image_charge}g, $g_{\rm dd}$ varies by a factor of three if the interface states are displaced in opposite directions. Finally, the variation in $g_{\rm dd}$ can be reduced even further by fabricating the gate stack in such a way that the charges in neighboring qubits are displaced in perpendicular directions, as in Fig. \ref{fig:image_charge}e. In this way, from Eq. \ref{eq:g_dd_ic}, the only dipole terms contributing to the coupling are the vertical ones, and therefore $g_{\rm dd}$ is unchanged (to first order) while tuning $V_t$.

\end{document}